\documentclass[12pt,reqno,a4paper]{article}
\usepackage{amsmath,amsfonts,amssymb,graphicx}
\numberwithin{equation}{section}

\newcommand{\nc}{\newcommand}
\nc{\rnc}{\renewcommand}

\nc{\be}{\begin{equation}}
\nc{\round}[1]{\left(#1\right)}
\nc{\squareb}[1]{\left[#1\right]}
\nc{\curly}[1]{\left\{#1\right\}}
\nc{\abs}[1]{\left|#1\right|}
\nc{\pointy}[1]{\left\langle #1 \right\rangle}
\nc{\lsum}[3]{\sum_{#1=#2}^{#3}}
\def\phat{p}
\def\th{\vartheta}
\def\half {\mbox{$\textstyle {1 \over 2}$}}

\def\Re{\mbox{Re}}
\def\Im{\mbox{Im}}
\nc{\one}{$1$\nobreakdash}
\nc{\two}{$2$\nobreakdash}
\rnc{\u}{$u$\nobreakdash}
\nc{\vm}{\boldsymbol{m}}
\nc{\vn}{\boldsymbol{n}}
\nc{\ve}{\boldsymbol{e}}
\nc{\vI}{\mathcal{I}}
\nc{\A}{\mathrm{A}}
\def\ln{\log}

\nc{\Wt}[6]{#1\!\!\left.\left(
\begin{matrix}
#5 & #4\\
#2 & #3
\end{matrix}
\right|#6\right)}

\nc{\Ktl}[4]{K
\left(\left.
\begin{matrix}
#3\\
#1
\end{matrix}
\ \;#2\right|#4\right)}

\nc{\Ktr}[4]{K
\left(\left.
#2 \ \;
\begin{matrix}
#3\\
#1
\end{matrix}
\right|#4\right)}

\nc{\Kt}[3]{K
\begin{pmatrix}
#2 &
\begin{matrix}
#3\\
#1
\end{matrix}
\end{pmatrix}}

\nc{\Db}{D_{\text{b}}}
\nc{\Ds}{D_{\text{s}}}
\nc{\Df}{D_{\text{f}}}
\nc{\Tb}{T_{\text{b}}}
\nc{\ks}{\kappa_{\text{s}}}
\nc{\kb}{\kappa_{\text{b}}}
\nc{\fsing}{f_{\text{sing}}}
\nc{\fsinglr}{\fsing^{\text{L},\text{R}}}
\nc{\Cb}{C_{\text{b}}}
\nc{\Cs}{C_{\text{s}}}
\nc{\als}{\alpha_{\text{s}}}
\nc{\fb}{f_{\text{b}}}
\nc{\fs}{f_{\text{s}}}

\nc{\xil}{\xi_{\text{L}}}
\nc{\xir}{\xi_{\text{R}}}
\nc{\al}{r}
\nc{\ar}{s}
\nc{\kslr}{\ks^{\text{L},\text{R}}}
\nc{\epsl}{\eps^{\text{L}}}
\nc{\epsr}{\eps^{\text{R}}}
\nc{\epslr}{\eps^{\text{L},\text{R}}}

\nc{\thf}{\vartheta_1}
\nc{\thfr}{\vartheta_4}

\nc{\lam}{\lambda}
\nc{\Lam}{\Lambda}
\nc{\Om}{\Omega}
\nc{\gam}{\gamma}
\nc{\eps}{\epsilon}
\nc{\veps}{\varepsilon}
\nc{\kap}{\kappa}
\nc{\m}{\mu}

\nc{\D}{\boldsymbol{D}}
\nc{\av}{\boldsymbol{a}}
\nc{\bv}{\boldsymbol{b}}
\nc{\ttt}{\boldsymbol{t}}
\nc{\I}{\boldsymbol{I}}

\nc{\sm}[1]{{\scriptstyle #1}}
\nc{\ssm}[1]{{\scriptscriptstyle #1}}
\nc{\spos}[2]{\makebox(0,0)[#1]{$\sm{#2}$}}
\nc{\sposb}[2]{\makebox(0,0)[#1]{$ #2 $}}
\nc{\dl}[3]{\put(#1,#2){\makebox(#3,0){\dotfill}}}
\rnc{\d}[2]{\put(#1,#2){\spos{}{\bullet}}}
\nc{\dd}[3]{\multiput(#1,#2)(0,1){#3}{\spos{}{\bullet}}}

\def\u{\epsilon}
\def\b{\beta}

\def\h{\vartheta}
\def\xR{\xi_R}
\def\xL{\xi_L}

\def\r2{\rho_2}
\def\({\biggl(}
\def\){\biggr)}

\def\l{\lambda}
\nc{\Mult}[2]{\genfrac{[}{]}{0pt}{}{#1}{#2}}
\nc{\qbin}[2]{\Mult{#1}{#2}}

\rnc{\title}[1]{{\Large\bf\mbox{}\\\medskip#1\bigskip\medskip\\}}
\rnc{\author}[1]{{\large #1\smallskip\\}}
\nc{\address}[1]{{\em #1\medskip\\}}

\begin{document}

\begin{center}

\title{Excited TBA Equations I:\\
Massive Tricritical Ising Model}

\author{
Paul A. Pearce,\!\!
\footnote{
P.Pearce@ms.unimelb.edu.au}\;
Leung Chim\!\!
\footnote{Current address: DSTO, Adelaide;
Leung.Chim@dsto.defence.gov.au}
}
\address{Department of Mathematics and Statistics\\
University of Melbourne,
Parkville, Victoria 3052, Australia}
\author{Changrim Ahn\!
\footnote{ahn@dante.ewha.ac.kr}}
\address{Department of Physics\\
Ewha Womans University, Seoul 120-750, Korea}


\begin{abstract}
\noindent
We consider the massive tricritical Ising model ${\cal M}(4,5)$ perturbed by
the thermal operator $\varphi_{1,3}$ in a cylindrical geometry and apply
integrable boundary conditions, labelled by the Kac labels $(r,s)$, 
that are natural
off-critical perturbations of known conformal boundary conditions. We 
derive massive
thermodynamic Bethe ansatz (TBA) equations for all excitations by 
solving, in the continuum
scaling limit, the TBA functional equation satisfied by the double-row 
transfer matrices
of the $A_4$ lattice model of Andrews, Baxter and Forrester (ABF) in 
Regime~III. The
complete classification of excitations, in terms of $(\vm,\vn)$ 
systems, is precisely the
same as at the conformal tricritical point. Our methods also apply on 
a torus but we first
consider $(r,s)$ boundaries on the cylinder because the classification of
states is simply related to fermionic
representations of single Virasoro characters 
$\chi_{r,s}(q)$. We study the TBA equations analytically and 
numerically to determine
the conformal UV and free particle IR spectra and the connecting massive flows. The TBA equations in
Regime IV and massless RG flows are studied in Part~II.
\end{abstract}
\end{center}

\section{Introduction}

Ever since their introduction~\cite{YangYang,Zam90,Zam91},
Thermodynamic Bethe Ansatz (TBA) equations have been an important tool in
the study of both massive and massless integrable quantum field 
theories. 
Although extensive studies have been carried out on scaling energies of 
vacuum or ground states only relatively few excited 
states~\cite{Martins,Fendley,Dorey,BLZ,Rav}
have proven amenable to TBA analysis and these are 
primarily restricted to massive
and diagonal scattering theories. So despite considerable successes, 
the application
of TBA methods has been hampered by inherent limitations. The primary 
obstacle is
that to date there is no systematic and unified derivation of
excited state TBA equations. Indeed, the current treatments of 
excited states are at
best ad hoc and fall well short of a complete analysis of all 
excitations. Here we
address these limitations and propose a systematic approach based on 
the lattice.
More specifically, we will show in a series of papers that both 
massive and massless
excited TBA equations can be systematically obtained by studying the continuum
scaling limit of the associated integrable lattice models. Perhaps the most
important input from the lattice approach is an insight into the analytic
structure of the excited state solutions of the TBA equations. Previously this
structure had to be guessed. In stark contrast, in the lattice approach, the
analyticity structure can be probed by direct numerical calculations 
on finite size
transfer matrices.

Although the methods developed in this paper are very general, for 
simplicity and
concreteness, we consider as a first example the massive
tricritical Ising model ${\cal M}(4,5)$ perturbed by the thermal operator
$\varphi_{1,3}$. Although this is a non-diagonal scattering theory 
and more complicated
from the viewpoint of integrable quantum field theory, this is the 
simplest case
beyond the Ising model and Lee-Yang theory for analysis by the 
lattice approach. The
integrable lattice model associated to the thermally perturbed 
tricritical Ising
model is the interacting hard square model or generalized hard 
hexagon model solved
by Baxter~\cite{Bax80,Bax82}. This model, with its ${\mathbb Z}_2$ 
sublattice symmetry,
is known to be in the universality class of the tricritical Ising model. More
generally, this model is the special case $L=4$ of the $A_L$ lattice models of
Andrews, Baxter and Forrester~\cite{ABF84} with the $L=3$ model being the
usual Ising model. Generically these $A_L$ models, with their ${\mathbb
Z}_2$ height
reversal symmetry, are in the universality class of multicritical Ising models.
Moreover, the continuum scaling limit of the
$A_L$ models realize~\cite{Huse84} the $\varphi_{1,3}$ thermal 
perturbation of the
$s\ell(2)$ unitary minimal models~\cite{BPZ84} and the ground states 
are described
by the
$A_{L-2}$ TBA equations of Zamolodchikov~\cite{Zam91}.

There have been many relevant studies of the tricritical hard square 
or $A_4$ lattice
model and the more general $A_L$ models from the lattice viewpoint. For the
$A_4$ model, the off-critical TBA functional equation for periodic boundary
conditions has been derived and solved~\cite{BaxP82,BaxP83} for the 
bulk properties
and correlation lengths. The off-critical TBA functional equations 
for the $A_L$
models were derived by Kl\"umper and 
Pearce~\cite{PearK91,KlumP91,KlumP92}. But only
the critical or ``conformal TBA" equations were derived and solved in 
the critical scaling
limit for the central charges and conformal weights. The very same 
off-critical TBA
functional equations for
$A_L$ models were subsequently derived~\cite{BPO96} in the presence 
of integrable
boundaries showing that the TBA functional equations are universal in 
the sense that
they are independent of the boundary conditions. A biproduct of introducing
boundaries is that the problem of classifying the excitations becomes 
much easier.
This is reflected in the fact that at criticality the cylinder partition
functions are given as linear forms in characters rather than the 
usual sesquilinear
form on the torus. Indeed, by a judicious choice of $(r,s)$ boundary 
conditions, the
cylinder partition function is just a single Virasoro character $\chi_{r,s}(q)$
and the complete classification of excitations~\cite{OPW97} in terms 
of $(\vm,\vn)$
systems~\cite{Melzer,Berk94} is related to a fermionic representation of the
character $\chi_{r,s}(q)$. This simplification enabled~\cite{OPW97} the complete analytic
calculation of the conformal cylinder partition functions of the 
$A_4$ model with
6 different conformal boundary conditions $(r,s)$ which are conjugate 
to the 6 primary
fields of the tricritical Ising conformal field theory. The 
generalization of these
results to the critical
$A_L$ lattice models with $(r,s)$ boundaries is currently in 
progress~\cite{PearW00}.

In 1998, Pearce and Nienhuis~\cite{PearN98} analysed the off-critical
continuum scaling limit of the TBA functional equations for the $A_L$ 
lattice models
to derive from the lattice the full set of massive and massless 
$A_{L-2}$ ground
state TBA equations conjectured by Zamolodchikov. The scale parameter 
$mR$ is simply
related to the scaling limit of lattice parameters by
\begin{equation}
\mu=\frac{mR}{4}=\lim_{N\to\infty,\, t\to 0} N t^\nu
\end{equation}
or more precisely
\begin{equation}
R=\lim_{N\to\infty,\, a\to 0} Na,\qquad m=\lim_{t\to 0,\, a\to 
0}\frac{4t^\nu}{a}
\end{equation}
where $a$ is the lattice spacing, $m$ is a mass, $R$ is the continuum length 
scale, $\nu=(L+1)/4$ is the correlation length exponent and $t=p^2$ is the
deviation from critical temperature variable with
$p$ the elliptic nome appearing in the Boltzmann weights of the $A_L$ 
models. 

Our primary goal in this series of papers is to extend the analysis of Pearce and
Nienhuis to all excitations of the $A_4$ model both in the massive and massless
regimes. This entails perturbing the analysis of O'Brien, Pearce and
Warnaar~\cite{OPW97} off criticality. To handle the problem of 
classification of all
the excitations it is easier to introduce boundaries and to work on the 
cylinder even
though a study of boundary properties is not our primary goal. Our 
immediate goal is
to study the flow of excitation energies from the UV ($R=0$) to the IR ($R\to
\infty$) limit. In the massive case we are thus able to compile the
conformal-massive dictionary that eluded Melzer~\cite{MelzerDict}. In 
the massless
case considered in paper~II~\cite{PCAII}, we follow the renormalization group
flow from the tricritical to the critical Ising model fixed points. This leads to a 
flow between the
characters of these theories.

In this paper we consider just the massive regime. The layout of the paper
is as follows. In Section~2 we define the $A_4$ lattice model. We 
then describe the
classification of excitations and discuss their unique labelling in
terms of quantum numbers. This classification is in fact identical to the
classification~\cite{OPW97} at the tricritical point. In Section~3 we 
present the
derivation of the off-critical massive TBA equations. The numerical solution of
these equations is presented in Section~4. Throughout we concentrate 
on the three $(r,s)$
boundary conditions with $s=1$ rather than presenting exhaustive 
results for the six
distinct $(r,s)$ boundary conditions. We believe these results are 
indicative of
what can be achieved. While the calculations are similar for the other boundary
conditions, we point out that in some cases there are further 
subtleties related to
the appearance of frozen zeros. We conclude with a general discussion.

\section{Lattice Approach}
\subsection{$A_4$ lattice model}
The $A_4$ RSOS lattice model is defined on a square lattice with 
spins or heights
$a=1,2,3,4$ restricted so that nearest neighbour heights differ by 
$\pm 1$. This
model corresponds to the special case $L=4$ of the $A_L$ RSOS models 
of Andrews,
Baxter and Forrester~\cite{ABF84} with spins $a=1,\ldots,L$ and 
Boltzmann weights
($\lambda=\pi/5$ for the $A_4$ model)
\begin{eqnarray}
\Wt{W}{a}{a\mp1}{a}{a\pm1}u&=&\frac{\thf(\lam-u)}{\thf(\lam)}\\
\Wt{W}{a\mp1}{a}{a\pm1}{a}u&=&
\round{\frac{\thf((a-1)\lam)\thf((a+1)\lam)}{\thf^2(a\lam)}}^{1/2}\,
\frac{\thf(u)}{\thf(\lam)}\\
\Wt{W}{a\pm1}{a}{a\pm1}{a}u&=&\frac{\thf(a\lam\pm u)}{\thf(a\lam)}.
\end{eqnarray}
Here $u$ is the spectral parameter, $\lambda=\pi/(L+1)$ is the 
crossing parameter and
$\thf(u)=\thf(u,\phat)$ is one of the standard elliptic theta 
functions as given in
Gradshteyn and Ryzhik~\cite{GR}
\begin{eqnarray}
\thf(u,\phat)&=&2\phat^{1/4}\sin u
\prod_{n=1}^\infty (1-2\phat^{2n}\cos 2u+\phat^{4n})(1-\phat^{2n})\\
\th_2(u,\phat)&=&2\phat^{1/4}\cos u
\prod_{n=1}^\infty (1+2\phat^{2n}\cos 2u+\phat^{4n})(1-\phat^{2n})\\
\th_3(u,\phat)&=&\prod_{n=1}^\infty (1+2\phat^{2n-1}\cos 2u+\phat^{2(2n-1)})
(1-\phat^{2n})\\
\thfr(u,\phat)&=&\prod_{n=1}^\infty (1-2\phat^{2n-1}\cos 2u+\phat^{2(2n-1)})
(1-\phat^{2n}).
\end{eqnarray}
Integrability derives from the fact that these local face weights satisfy the
Yang-Baxter equation.

The elliptic nome $p$ is a temperature-like variable. The $A_L$ 
lattice models are
critical for $p=0$ and off-critical for $p\ne 0$ so we introduce the 
deviation from
critical temperature variable
$t=p^2$. There are four  distinct off-critical physical regimes 
depending on the sign of
$u$ and $t$:
\begin{eqnarray}
\begin{array}{lcr}
\text{Regime I:}&\qquad   -\pi/2+\lambda\le u \le 0,\qquad& -1<t<0\phantom{.}\\
\text{Regime II:} & -\pi/2+\lambda\le u \le 0,& 0<t<1\phantom{.}\\
\text{Regime III:}& 0\le u \le \lambda,& 0<t<1\phantom{.}\\
\text{Regime IV:}&  0\le u \le \lambda, &  -1<t<0.
\end{array}
\end{eqnarray}
It is convenient to express the nome
$\phat$ in terms of a real parameter $\veps>0$ by
\begin{equation}
\phat=\begin{cases} e^{-\pi\veps}, & \text{Regimes II and III}\\ ie^{-\pi\veps},
& \text{Regimes I and IV}
\end{cases}
\label{nome}
\end{equation}
so that $|p|\le 1$ and
\begin{equation}
t=\phat^2=\pm\exp(-2\pi\veps).
\end{equation}
In particular, the elliptic
$\thf$ functions satisfy the quasiperiodicity  properties
\begin{gather}
\thf(u+\pi,\phat)=-\thf(u,\phat)\\
\thf(u-i\ln \phat,\phat)=-\phat^{-1}e^{-2iu}\thf(u,\phat).
\end{gather}
Regimes III and IV are of interest in this series of papers since they are
associated, in the continuum scaling limit, with the massive and 
massless thermal
perturbations of the unitary minimal models respectively. In Regime~III, considered in
this paper, $p$ is real whereas in Regime~IV $p$ is pure imaginary. 
Regimes I and II relate
to ${\mathbb Z}_{L-1}$ parafermions and so are not considered here.

Although we will not use it in this paper, an alternative formulation 
of the $A_4$
model is the particle or $T_2$ tadpole representation as shown in Figure~1. This 
formulation is obtained
by folding the $A_4$ diagram and identifying the states related by 
the ${\mathbb Z}_2$
height reversal symmetry. More specifically, we can identify the 
states $a=1,4$ with
$\mu=1$ and regard this as indicating the presence of a particle or 
an occupied site
and we can identify the states $a=2,3$ with $\mu=0$ and regard this 
as indicating the
absence of a particle or a vacant site. Once we fix the sublattice of 
the square
lattice which has odd heights, the identification of height and 
particle states is a
one-to-one correspondence.  The adjacency constraint on the heights 
of the $A_4$
model translates into the exclusion of simultaneous occupancy of 
adjacent sites by
particles. In this way the particle representation is seen to be 
equivalent to a
model of interacting hard squares on the square lattice. The
Boltzmann weights of this hard square model are simply given by 
replacing the heights
$a=1,2,3,4$ by the corresponding particle occupation numbers $\mu=0,1$.

\begin{figure}[tbh]
\hskip 2.75truein
\includegraphics[width=.15\linewidth]{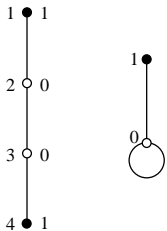}
    \caption{$A_4$ and $T_2$ graphs of allowed neighboring states for the $A_4$ lattice model and its
particle (hard square) representation.}\label{HSqTadpole}     
\end{figure}

The isotropic phase diagram of the $A_4$ or interacting hard square model is shown in Figure~2
alongside the corresponding renormalization group flow in the continuum scaling limit about the
tricritical point. The continuum scaling limits in Regimes~III and IV give rise to the massive and
massless flows respectively where the perturbation parameter is $mR$ or $\mu$.

\begin{figure}[htb]
\hskip 0.75truein
\includegraphics[width=.75\linewidth]{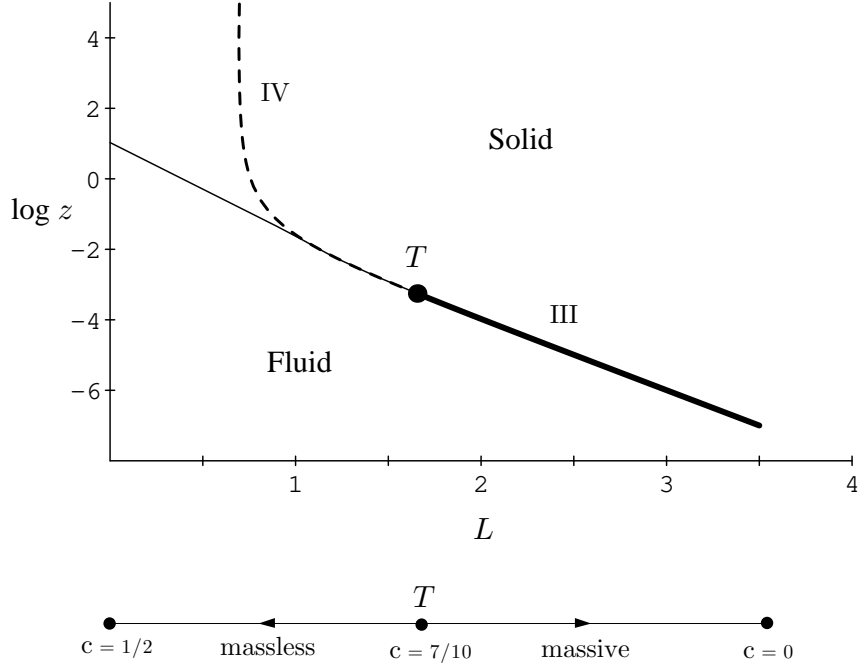}
    \caption{Isotropic phase diagram of the $A_4$ or interacting hard square model showing the phase
boundary (solid and thick solid curves) between the fluid and square ordered solid phases.  The
fugacity of the particles is
$z$ and
$L$ is the attraction between particles. The tricritical point
$T$ separates the line of Ising critical points (solid curve) from the first-order line (thick solid
curve, Regime~III). The model is exactly solvable on the first-order line and its analytic continuation
(dashed line, Regime~IV). The exact solution curve  and Ising critical line are tangential at $T$. The
renormalization group flow in the continuum scaling limit about the tricritical point $T$ is shown
below for comparison. The central charge $c$ is indicated at the tricritical,
critical and trivial fixed points.}\label{HSqPhDiag}     
\end{figure}

\subsection{Double row transfer matrices}
To ensure integrability of the $A_4$ model in the presence of a 
boundary~\cite{BPO96}
we need commuting double row transfer matrices and triangle boundary 
weights which
satisfy the boundary Yang-Baxter equation. The triangle weights for the $(r,s)$
boundary condition with $1\le r\le 3$ and $1\le s\le 4$ are given by
\begin{equation}
\begin{split}
\Ktl{r}{r\pm 1}{r}{u,\xi_L}&=\round{\frac{\th_1((r\pm
1)\lam)}{\th_1(r\lam)}}^{\frac{1}{2}}
\dfrac{\th_1(u\pm\xi_L)\th_1(u\mp r\lam\mp \xi_L)}
{\th_1^{2}(\lam)}\\
\Ktr{s}{s\pm 1}{s}{u,\xi_R}&=
\round{\frac{\th_1((s\pm 1)\lam)}{\th_1(s\lam)}}^{\frac{1}{2}}
\dfrac{\th_4(u\pm\xi_R)\th_4(u\mp s\lam\mp \xi_R)}
{\th_4^{2}(\lam)}.
\end{split}
\label{Kdef}
\end{equation}
The parameters $\xi_L,\xi_R$ are arbitrary and can be taken to be complex,
however here we restrict them to the real interval
$(\lam/2,\lam)$. To obtain conformal boundary
conditions at the isotropic tricritical point $u=\lam/2$, $t=p=0$ we should
choose~\cite{BP00} $\xi_L=\xi_R=\lam/2$. Integrability in the presence of these
boundaries derives from the fact that these boundary triangle weights 
satisfy the
boundary Yang-Baxter equation.

The face and triangle boundary weights are used to construct~\cite{BPO96} a
family of commuting double row transfer matrices $\D(u)$.  For a 
lattice of width
$N$, the  entries of $\D(u)$ are given
diagrammatically by
\setlength{\unitlength}{14mm}
\begin{equation}
\D(u)_{\av,\bv}
=\sum_{c_{0},\dots,c_{N}}
\raisebox{-1.4\unitlength}[1.3\unitlength][
1.1\unitlength]{\begin{picture}(6.4,2.4)(0.4,0.1)
\multiput(0.5,0.5)(6,0){2}{\line(0,1){2}}
\multiput(1,0.5)(1,0){3}{\line(0,1){2}}
\multiput(5,0.5)(1,0){2}{\line(0,1){2}}
\multiput(1,0.5)(0,1){3}{\line(1,0){5}}
\put(1,1.5){\line(-1,2){0.5}}\put(1,1.5){\line(-1,-2){0.5}}
\put(6,1.5){\line(1,2){0.5}}\put(6,1.5){\line(1,-2){0.5}}
\put(0.5,0.45){\spos{t}{\al}}\put(1,0.45){\spos{t}{\al}}
\put(2,0.45){\spos{t}{a_1}}\put(3,0.45){\spos{t}{a_2}}
\put(5,0.45){\spos{t}{a_{N-1}}}\put(6,0.45){\spos{t}{\ar}}
\put(6.5,0.45){\spos{t}{\ar}}
\put(0.5,2.6){\spos{b}{\al}}\put(1,2.6){\spos{b}{\al}}
\put(2,2.6){\spos{b}{b_{1}}}\put(3,2.6){\spos{b}{b_2}}
\put(5,2.6){\spos{b}{b_{N-1}}}\put(6,2.6){\spos{b}{\ar}}
\put(6.5,2.6){\spos{b}{\ar}}
\put(1.05,1.45){\spos{tl}{c_0}}\put(2.05,1.45){\spos{tl}{c_1}}
\put(3.05,1.45){\spos{tl}{c_2}}\put(4.99,1.45){\spos{tr}{c_{N-1}}}
\put(5.99,1.45){\spos{tr}{c_{N}}}
\multiput(1.5,1)(1,0){2}{\sposb{}{u}}\put(5.5,1){\sposb{}{u}}
\multiput(1.5,2)(1,0){2}{\sposb{}{\lam\!-\!u}}
\put(5.5,2){\sposb{}{\lam\!-\!u}}
\put(0.71,1.5){\sposb{}{\lam\!-\!u\ \ \ }}\put(6.29,1.5){\sposb{}{u}}
\multiput(0.5,0.5)(0,2){2}{\makebox(0.5,0){\dotfill}}
\multiput(6,0.5)(0,2){2}{\makebox(0.5,0){\dotfill}}
\end{picture}}
\label{RTMdef}
\end{equation}
It is further convenient to define the normalized transfer matrix
\begin{equation}
\label{e:t}
\ttt(u) = S_{r,s}(u)S(u)
\Bigl[i\,\frac{\h_1(u+2\l,p)\h_1(\l,p)}{\h_1(u+3\l,p)\h_1(u+\l,p)}\Bigr]^{2N}
\D(u)
\end{equation}
where
\begin{equation}
S(u)=\frac{\h_1(2u-\l,p)^2}{\h_1(2u-3\l,p)\h_1(2u+\l,p)}
\end{equation}
and
\begin{equation}
S_{r,s}(u) 
  = (-1)^{s}h_r(u-\xL)h_{-r}(u+\xL)\bar{h}_{s}(u-\xR)\bar{h}_{-s}(u+\xR)
\end{equation}
with
\begin{align}
h_r(u)& = \frac{\h_1(\l,p)\h_1(u+(3-r)\l,p)\h_1(u+(1-r)\l,p)}
{\h_1(u,p)\h_1(u-\l,p)\h_1(u+2\l,p)}\\
\bar{h}_s(u)& = \frac{\h_4(\l,p)\h_4(u+(3-s)\l,p)\h_4(u+(1-s)\l,p)}
{\h_4(u,p)\h_4(u-\l,p)\h_4(u+2\l,p)}.
\end{align}
It can then be shown~\cite{BPO96} that the normalized transfer matrix 
satisfies the
universal TBA functional equation
\begin{equation}
\label{e:functional}
\ttt(u)\ttt(u+\l) = \I + \ttt(u+3\l)
\end{equation}
independent of the boundary condition $(r,s)$. This is precisely the same TBA
functional equation that holds in the periodic case~\cite{BaxP82}. Since the
transfer matrices commute this functional equations also holds for each
eigenvalue $t(u)$ of $\ttt(u)$.

The TBA functional equations will be solved for the finite-size 
corrections to the
eigenvalues $D(u)$ of the double row transfer matrices $\D(u)$. In the scaling
limit, the finite-size corrections to the eigenvalues $D(u)$ are related
to the excitation energies $E(R)$ of the associated perturbed 
conformal field theory
by
\begin{equation}
-\half \log D(u)= N
f_{\text{bulk}}(u)+b_{r,s}(u)+\frac{R\sin\vartheta}{N}\,E(R)+o(\frac{1}{N})
\end{equation}
where $f_{\text{bulk}}(u)$ is the bulk free energy, $b_{r,s}(u)$ is 
the boundary
free energy and the anisotropy angle is given by
\begin{equation}
\vartheta=
\begin{cases} (L+1)u,&\text{Regimes III and IV}\\
-\displaystyle{\frac{2(L+1)u}{L-1}},&\text{Regimes I and II.}
\end{cases}
\end{equation}
Depending on the boundary
conditions, the boundary free energy $b_{r,s}(u)$ may contain an interfacial
free energy contribution. The bulk and boundary free energies can be calculated~\cite{Stroganov,
Bax82Inv,OBrienP97} by the inversion relation method. Despite the appearance of
$1/N$ corrections, the system is not in general  conformally invariant. The system is 
conformal however
at critical points which can occur in the  ultraviolet
($R\to 0$) and infrared ($R\to\infty$) limits with
\begin{equation}
\frac{R E(R)}{2\pi}\to -\frac{c}{24}+\Delta_{r,s}+n,\qquad n\in
{\mathbb N}
\end{equation}
where $c$ is the central charge of the appropriate conformal field 
theory, $\Delta_{r,s}$
are the related conformal weights and
$n=0,1,\ldots$ labels the tower of descendants. The largest 
eigenvalue occurs for
the vacuum or ground state with the boundary condition $(r,s)=(1,1)$. 
In this case
$\Delta_{1,1}=0$ and $n=0$. The massive
$R\to\infty$ scaling limit in Regime~III, however, is trivial in the sense
that for this ground state $RE(R)\to 0$ corresponding to $c=0$ and 
the scattering
of free massive particles.

\subsection{Classification: $(\vm,\vn)$ systems and quantum numbers}

TBA functional equations admit infinite families of solutions for the
eigenvalues $t(u)$. The analyticity properties are therefore crucial in
selecting out the required solutions.  The transfer matrix 
eigenvalues $D(u)$ are
entire functions of $u$ and are characterised (up to an overall 
constant) by their
zeros in the complex $u$ plane. It is precisely at these zeros that 
$\log t(u)$ is
non-analytic but analyticity of $\log t(u)$ is required to solve the 
TBA functional
equations by Fourier series. From quasiperiodicity, the matrix
$\ttt(u)$ and eigenvalues
$t(u)$ are doubly periodic. It follows
that the eigenvalues $t(u)$ are doubly periodic meromorphic functions.
It is convenient to fix the period rectangles as
\begin{equation}
\text{period rectangle}=
\begin{cases}
(-\frac{\lambda}{2},\frac{9\lambda}{2})\times
(-\frac{\pi i\veps}{2},\frac{\pi i\veps}{2}),
&\text{Regime~III}\\
(-\frac{\lambda}{2},\frac{9\lambda}{2})\times
(-\pi i\veps,\pi i\veps),&\text{Regime~IV}
\end{cases}
\end{equation}
so we then only need to consider the analyticity inside these period rectangles.
In Regime~IV there is an additional symmetry within the period rectangle
\begin{equation}
t(u\pm\pi/2+\pi i\veps)=t(u)
\end{equation}
so we can restrict ourselves further to the rectangle
$(-\frac{\lambda}{2},2\lambda)\times (-\pi i\veps,\pi i\veps)$.
The normalization factors relating $D(u)$ to $t(u)$ only introduce 
extra zeros and
poles on the real axis. Since $\D(u)$ is real symmetric for real $u$, that is
$\D(u)=\D(u)^T$, it follows that, for any eigenvalue $D(u)$ or $t(u)$, the
distribution of zeros in the upper and lower half planes are identical and
simply related by complex conjugation. It is therefore sufficient to 
classify the
eigenvalues by the patterns of zeros in the upper half period rectangle.

It turns out that in Regime~III the pattern of zeros inside the periodic
rectangle is qualitatively the same as in the critical case. This was 
observed by
direct numerical diagonalization of a sequence of finite-size transfer matrices
approaching the scaling limit $N\to\infty$, $t\to 0$ for modest sizes 
of $N$.  As a
consequence, in Regime~III we can use the known 
classification~\cite{OPW97} of eigenvalues
at the tricritical point ($R=0$) in terms of $(\vm,\vn)$ systems and this
classification will apply for any
$R$ in the range $0\le R<\infty$. This simplifying feature
does not hold in the massless Regime~IV where in stark contrast some 
of the patterns of
zeros qualitatively change under the flow as $R$ increases~\cite{PCAII}.
Since the classification of excitations in Regime~III of interest 
here is the same as
at the tricritical point we summarise the salient features of this 
classification
here. We limit discussion to the boundary condition $(r,s)=(1,1)$. 
The other cases
are similar~\cite{OPW97} although in some cases it is necessary to 
introduce two
$(\vm,\vn)$ systems for a given $(r,s)$ boundary condition.

\begin{figure}[p]
\begin{center}
\includegraphics[width=.6\linewidth]{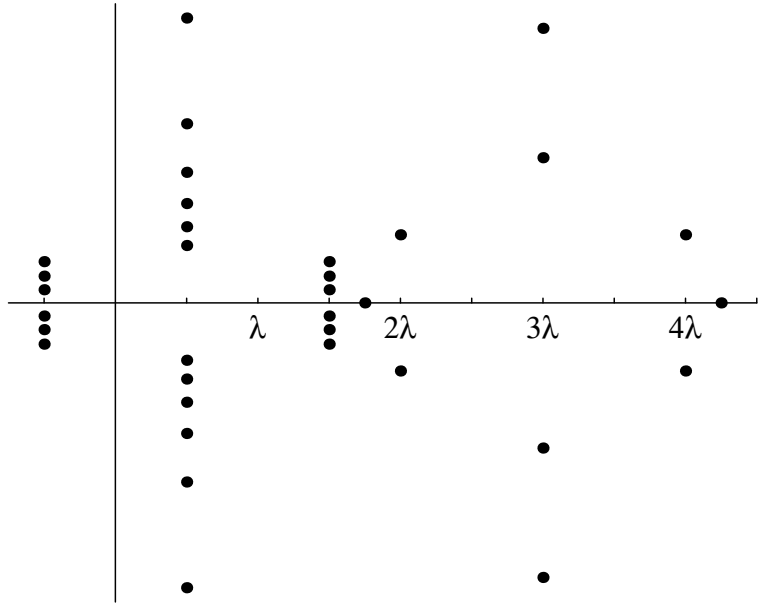}
\end{center}
\caption{Zeros within a period rectangle in
the complex $u$-plane of the largest eigenvalue of $D(u)$ with string 
content $m_1=6$,
$n_1=3$, $m_2=2$, $n_2=1$.}
\label{zpat11a}
\end{figure}

\begin{figure}[p]
\begin{center}
\includegraphics[width=.6\linewidth]{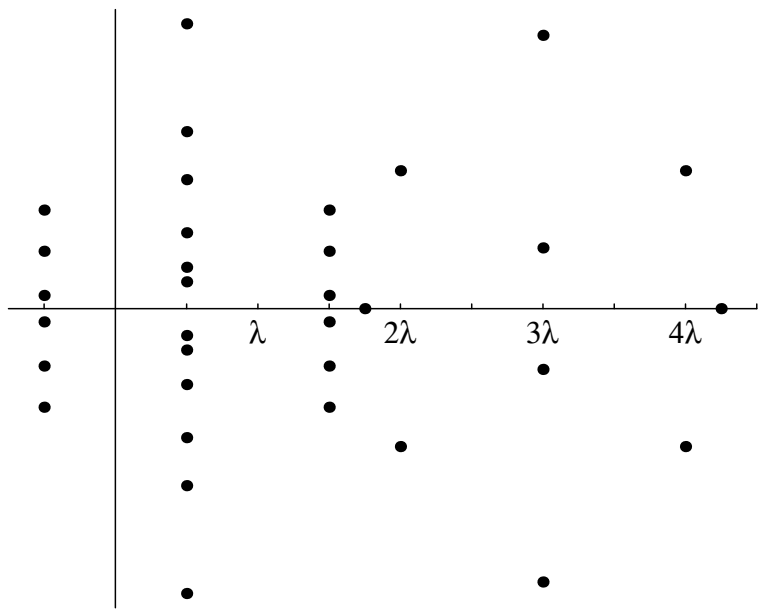}
\end{center}
\caption{Zeros within a period rectangle in the complex $u$-plane of another typical eigenvalue 
of $D(u)$ with string content $m_1=6$,
$n_1=3$, $m_2=2$, $n_2=1$. This pattern of zeros is obtained by permuting the ordering of the
1-strings and 2-strings within each strip of Figure~3.}
\label{zpat11b}
\end{figure}

The analyticity properties of $D(u)$ are relevant in the two analyticity strips
\begin{equation}
-\frac{\lambda}{2}<\Re(u)<\frac{3\lambda}{2},\qquad 2\lambda<\Re(u)<4\lambda
\end{equation}
We refer to these as strip~1 and 2 respectively and label them by 
$i=1,2$. Since
the Boltzmann weights are real and positive for $0<u<\lambda$, 
strip~1 is referred to as the physical analyticity strip. From direct numerical
diagonalization of
$\D(u)$ with the
$(1,1)$ boundary condition we observe that, apart from a pair of 
zeros on the real
axis at
$u=\lam+\xi$  and
$u=5\lam-\xi$ induced by the left boundary triangle weight, each eigenvalue has
zeros on the lines
$\Re(u)=-\lambda/2,\, \lambda/2,\, 3\lambda/2,\, 2\lambda,\, 4\lambda$
corresponding to the edges and center lines of the two analyticity strips.
Specifically, \one-strings and \two-strings occur in strip~1 and 2.  A
\one-string is  given by a single zero $u_{j}$ in the center of a 
strip such that
\begin{equation}
\Re(u_{j})=\begin{cases}
\lam/2, & \text{strip 1}\\
3\lam, & \text{strip 2.}
\end{cases}
\end{equation}
A \two-string is a pair of zeros $(u_{j},u'_{j})$ on the edge of a 
strip with equal
imaginary part and
\begin{equation}
(\Re(u_{j}),\Re(u'_{j}))=\begin{cases}
(-\lam/2,3\lam/2), & \text{strip 1}\\
(2\lam,4\lam), & \text{strip 2.}
\end{cases}
\label{twostrings}
\end{equation}
Distributions of zeros for two typical eigenvalues of $D(u)$ for
$N=16$ are depicted in Figures~\ref{zpat11a}
and~\ref{zpat11b}.
\nobreak
We note that for finite $N$ the \two-strings do not fall
precisely on the lines given in equation~\eqref{twostrings}, but that
this deviation decreases exponentially, that is as $|\delta|^N$ with 
$|\delta|<1$,
as $N$ increases. These patterns are consistent with the crossing 
$\D(u)=\D(\lam-u)$
and transpose $\D(u)=\D(u)^T$ symmetry of the double row transfer matrix.

Given an eigenvalue, we denote the number of strings in the upper half period
rectangle as follows:
\begin{equation}
\begin{split}
m_i&=\text{number of \one-strings in strip $i=1,2$}\\
n_i&=\text{number of \two-strings in strip $i=1,2$}.
\end{split}
\label{stringdefs}
\end{equation}
The relations~\cite{OPW97} between these numbers determining the 
string content take
the form of an $(\vm,\vn)$\nobreakdash-system~\cite{Melzer,Berk94}
\begin{equation}
\vm + \vn = \frac{1}{2}(N\ve_{1}+\vI\vm)
\label{mn11}
\end{equation}
where $\vm=(m_{1},m_{2})$, $\vn=(n_{1},n_{2})$, $\ve_{1}=(1,0)$, and
$\vI$ is the $\A_{2}$ incidence matrix with entries
$\vI_{j,k}=\delta_{|j-k|,1}$. Clearly here we require that $m_1, m_2$ 
and $N$ are
even. For the leading excitations $m_1, m_2, n_2$ are finite but 
$n_1\sim N/2$ as
$N\to\infty$. Indeed, the vacuum or ground state is given by 
$m_1=m_2=n_2=0$ and
$n_1=N/2$.

For each system size $N$, there are many eigenvalues with the
same string content $(\vm,\vn)$.  These eigenvalues are distinguished by
the relative vertical orderings of the $1$ and $2$-strings within the period rectangle 
along each strip.  Denoting the imaginary parts of
the \one-strings in strip~$i$ by
$0<v_{1}^{(i)}<\dots<v_{m_{i}}^{(i)}$, and the imaginary
parts of the \two-strings in strip~$i$ by
$0<w_{1}^{(i)}<\dots<w_{n_{i}}^{(i)}$, we see that in
Figure~\ref{zpat11a}
\begin{equation}
\begin{gathered}
0<w_{1}^{(1)}<w_{2}^{(1)}<w_{3}^{(1)}
<v_{1}^{(1)}<v_{2}^{(1)}<\dots<v_{6}^{(1)}\\
0<w_{1}^{(2)}<v_{1}^{(2)}<v_{2}^{(2)}
\end{gathered}
\end{equation}
whereas in Figure~\ref{zpat11b}, which has the same string content,
\begin{equation}
\begin{gathered}
0<w_{1}^{(1)}<v_{1}^{(1)}<v_{2}^{(1)}<w_{2}^{(1)}<v_{3}^{(1)}
<w_{3}^{(1)}<v_{4}^{(1)}<v_{5}^{(1)}<v_{6}^{(1)}\\
0<v_{1}^{(2)}<w_{1}^{(2)}<v_{2}^{(2)}.
\end{gathered}
\end{equation}
Notice that the 1-strings $v_j^{(i)}$ and 2-strings $w_j^{(i)}$ labelled by $j=1$ are
closest to  the real axis.
Clearly, the total number of possible orderings for a given string 
content $(\vm,\vn)$
is $\binom{m_{1}+n_{1}}{m_{1}}\binom{m_{2}+n_{2}}{m_{2}}$.  Summing
over all allowed string contents $(\vm,\vn)$ using~\eqref{mn11} then
gives
\begin{equation}
\sum_{(\vm,\vn)}
\binom{m_{1}+n_{1}}{m_{1}}\binom{m_{2}+n_{2}}{m_{2}}
=A^{N}_{1,1}.
\label{count}
\end{equation}
This is indeed the correct number
of eigenvalues as given by the dimension of the double row transfer matrix, 
which is $A^{N}_{1,1}$, the number of
$N$-step paths from 1 to 1, where $A$ is the $A_4$ adjacency matrix.

In the scaling limit
\begin{equation}\label{e:scaling}
\mu=\frac{mR}{4}=\lim_{N\to\infty,\;t\to 0} N\,t^{5/4}
\end{equation}
the positions of the 1- and 2-strings grow logarithmically as
\begin{equation}
v^{(i)}_{j},w^{(i)}_{j}\,\sim\, -\frac{1}{4}\log t+\mbox{const}
\,=\,\frac{1}{5}\log N-\frac{1}{5}\log\mu+\mbox{const}.
\end{equation}
More specifically, we define the scaled locations of the strings in 
strips $i=1,2$ as
\begin{equation}
\begin{split}
\beta^{(i)}_{j}&=\lim_{N\to\infty,\,t\to 0}\left(5v^{(i)}_{j}+\frac{5}{4}\ln
t\right),\qquad j=1,2,\dots, m_1\\
\gamma^{(i)}_{k}&=\lim_{N\to\infty,\,t\to 0}\left(5w^{(i)}_{k}+\frac{5}{4}\ln
t\right),\qquad k=1,2,\dots, m_2
\end{split}
\label{yz}
\end{equation}

An excitation with string content $(\vm,\vn)$ is uniquely labelled by a set of
quantum numbers
\begin{equation}
I=(I^{(1)}|I^{(2)})=(I_1^{(1)},I_2^{(1)},\ldots,I_{m_1}^{(1)}\;|\;I_1^{(2)},I_2^{(2)},
\ldots,I_{m_2}^{(2)})
\end{equation}
where the integers $I_j^{(i)}\in{\mathbb N}$ with $i=1,2$ give the number of
2-strings whose imaginary parts $w_k^{(i)}$ are greater than that of the given 1-string $v_j^{(i)}$.
Clearly, the quantum numbers $I^{(j)}_{k}$ satisfy
\begin{equation}
n_{j}\ge I^{(j)}_{1}\ge I^{(j)}_{2}\ge\dots\ge I^{(j)}_{m_{j}}\ge 
0\qquad i=1,2.
\label{iranges}
\end{equation}
Conversely, given the quantum numbers we can read off the values of 
$m_1$ and $m_2$ and
then $n_1$ and $n_2$ are uniquely determined by the $(\vm,\vn)$ 
system. For given string
content
$(\vm,\vn)$, the lowest excitation occurs when all of the 1-strings 
are further out from
the real axis than all of the 2-strings. In this case  all of the 
quantum numbers vanish
$I_j^{(i)}=0$. Bringing the location of a 1-string closer to the real axis by
interchanging the location of the 1-string with a 2-string increments 
its quantum number by
one unit and increases the energy. At the tricritical point where 
$R=0$ these jumps in
energy are quantised in a tower with energy levels given by~\cite{OPW97}
\begin{equation}
E=\lim_{R\to 0}\frac{RE(R)}{2\pi}=-\frac{c}{24}
+\frac{1}{4}\vm C \vm
+\sum_{i=1}^{2}\sum_{j=1}^{m_{i}}
I^{(i)}_j.
\label{sameaslater}
\end{equation}
Here $C$ is the $A_2$ Cartan matrix and the central charge is $c=7/10$.

More succinctly, the generating function for the finite-size spectra of a
cylindrical lattice built by $M$ applications of the double row 
transfer matrix with
$(1,1)$ boundary condition is given by a finitized Virasoro 
character~\cite{Melzer,Berk94}
\begin{equation}
\label{FinVirChar}
Z_{(1,1)}^{(N)}(q)=\sum_E q^E=q^{-c/24}\sum_{(\vm,\vn)}
q^{\frac{1}{4} \vm C \vm} \qbin{m_1+n_1}{m_1}\qbin{m_2+n_2}{m_2}
=\chi_{1,1}^{(N)}(q)
\end{equation}
where the sum is over the finite $(\vm,\vn)$ system, $q$ is the 
modular parameter
\begin{equation}
q=e^{-2\pi\frac{M}{N}\sin\vartheta}
\end{equation}
and $M/N$ is the aspect ratio of the lattice. In the isotropic case when
$u=\lambda/2$, the anistropy angle $\vartheta=\pi/2$ and the geometric factor
$\sin\vartheta=1$.
The $q$-binomial or Gaussian polynomial is defined by
\begin{equation}
\qbin{m+n}{m}=
\sum_{I_1=0}^n \sum_{I_2=0}^{I_1}\cdots \sum_{I_m=0}^{I_{m-1}} q^{I_1+\ldots+I_m}
=\begin{cases}
\dfrac{(q)_{m+n}}{(q)_{m}(q)_{n}}, &\quad m,n\ge 0 \\
0, &\quad \text{otherwise}\\
\end{cases}
\end{equation}
with the $q$-factorials $(q)_{m}=(1-q)\cdots(1-q^{m})$ for $m\ge 1$ 
and $(q)_{0}=1$.
In the limit $q\to 1$ the $q$-binomials reduce to the usual binomial 
coefficients
and the partition function just counts the number of states as in 
(\ref{count}). Note also that
\begin{equation}
\lim_{n\to\infty}\qbin{m+n}{m}=\qbin{\infty}{m}=\frac{1}{(q)_m}.
\end{equation}
After using the $(\vm,\vn)$ system to eliminate $n_1$ and $n_2$, the finitized
character gives the fermionic representation of the usual Virasoro 
character in the
limit $N\to\infty$
\begin{equation}
\lim_{N\to\infty}\chi_{1,1}^{(N)}(q)=
q^{-c/24}\sum_{m_1,\,m_2\, \text{even}}
\frac{q^{\frac{1}{4} \vm C \vm}}{(q)_{m_1}}\, \qbin{\frac{1}{2}m_1}{m_2}
=\chi_{1,1}(q).
\end{equation}

More generally, as explained in \cite{OPW97}, the
finitized partition functions for $(r,s)$ boundary conditions at 
$R=0$ are given in
terms of finitized characters
\begin{equation}
Z_{(r,s)}^{(N)}(q)=q^{-\frac{c}{24}+\Delta_{r,s}-\frac{1}{4}(s-r)(s-r-1)}
\sum_{(\vm,\vn)}
q^{\frac{1}{4} \vm C \vm-\frac{1}{2}m_{s-1}} \prod_{i=1,2}\qbin{m_i+n_i}{m_i}
=\chi_{r,s}^{(N)}(q)
\end{equation}
where the $(\vm,\vn)$ system for the $(r,s)$ boundary condition is
\begin{equation}
\vm + \vn = \frac{1}{2}(N\ve_{1}+\vI\vm+\ve_{s-1}+\ve_{4-r}).
\label{mnrs}
\end{equation}
Hence in the limit $N\to \infty$ we recover the known conformal 
cylinder partition
functions in terms of Virasoro characters
\begin{equation}
Z_{r,s}(q)=\lim_{N\to\infty}\chi_{r,s}^{(N)}(q)=\chi_{r,s}(q).
\end{equation}

\section{TBA Equations in Regime~III}

\subsection{Sector $(r,s)=(1,1)$}

In this section we derive the TBA equations for the $(r,s)=(1,1)$ boundary by
solving the TBA functional equations in the scaling limit for even 
$N$. We follow
closely the derivations in \cite{OPW97} and \cite{PearN98}. 
The derivation for other boundary conditions is similar. 
We begin by
factorizing $t(u)$ for large $N$ as
\begin{equation}
t(u) = f(u)g(u)l(u)
\end{equation}
where $f(u)$ accounts for the bulk order-$N$ behaviour, $g(u)$ the
order-$1$ boundary contributions and $l(u)$ is the order-$1/N$
finite-size correction. We will solve for $f(u)$, $g(u)$ and then $l(u)$
sequentially.

For the order-$N$ behaviour the second term on the RHS of the TBA
functional equation \eqref{e:functional} can be neglected giving the inversion
relation
\begin{equation}\label{e:bulkf}
f(u)f(u+\l) = 1.
\end{equation}
The prefactor in \eqref{e:t} induces poles of order $2N$ at
$u=2\l$ and $u=4\l$, and a zero of order $2N$ at $u=3\l$.
The required solution with this analyticity is
\begin{equation}
\label{e:f3}
f(u) = \begin{cases}
1,& \text{$-\frac{\l}{2}<\Re(u)<\frac{3\l}{2}$}\\
\displaystyle{\Bigl[i\frac{\h_2(\frac{5u}{2},t^{5/4})}
{\h_1(\frac{5u}{2},t^{5/4})}\Bigr]^{2N}},&
\text{\;\,$\frac{3\l}{2}<\Re(u)<\frac{9\l}{2}$}.
\end{cases}
\end{equation}

Similarly to the critical case, putting this solution into the TBA functional
equations implies the order-1 functional equations for $g(u)$
\begin{equation}
g(u)g(u+\l) = \begin{cases}
1,& \text{$-\frac{\l}{2}<\Re(u)<\frac{3\l}{2}$}\\
1+g(u-2\l),& \text{$\;\,\frac{3\l}{2}<\Re(u)<\frac{9\l}{2}$}.
\end{cases}
\end{equation}
To solve for $g(u)$ we need to take into account the order-1 zeros and poles
introduced by the order-1 prefactor in \eqref{e:t}
\begin{equation}
S_{1,1}(u) = \frac{\h_1(\l,p)^2\h_4(\l,p)^2}{\h_1(u-\xL-\l,p)\h_1(u+\xL,p)
\h_4(u-\xR-\l,p)\h_4(u+\xR,p)}.
\end{equation}
The order-1 zeros of $D(u)$ cancel exactly the poles of $S_{1,1}(u)$.
However $S(u)$ introduces poles at $u=-\frac{\l}{2}+i\frac{\rho\pi\u}{2},
\frac{3\l}{2}+i\frac{\rho\pi\u}{2}, 2\l+i\frac{\rho\pi\u}{2},
4\l+i\frac{\rho\pi\u}{2}$, and
double zeros at $u=\frac{\l}{2}+i\frac{\rho\pi\u}{2},
3\l+i\frac{\rho\pi\u}{2}$ where $\rho=0,\pm1$.
Thus the solution in strip~1 is given by
\begin{equation}
g(u) = - \Bigl[\frac{\h_1(5(u-\l/2)/2,t^{5/8})}
{\h_2(5(u-\l/2)/2,t^{5/8})}\Bigr]^2.
\end{equation}

The solution for $g(u)$ in strip~2 is more involved and requires
solving the functional relation for $\log g(u)$ by Fourier series. To proceed
we fix lines of constant real part
in the centers of each of the two strips in the $u$-plane and a real 
coordinate $x$
by
\begin{equation}
u=\begin{cases}
\frac{\l}{2}+\frac{ix}{5},&\text{strip~1}\cr
3\l+\frac{ix}{5},&\text{strip~2}.
\end{cases}
\end{equation}
It is then natural to define generically for the functions 
$h=t,f,g,l$ the
notations
\begin{subequations}
\begin{align}
h_1(x)& = h(\frac{\l}{2}+\frac{ix}{5}), \quad \;\text{$|\Im(x)|<\pi$}\\
h_2(x)& = h(3\l+\frac{ix}{5}), \quad \text{$|\Im(x)|<\pi$}\\[5pt]
H_1(x) =& 1+h_1(x), \quad H_2(x) = 1+h_2(x).
\end{align}
\end{subequations}
In the variable $x$, the functional relations become
\begin{subequations}\label{e:g3}
\begin{align}
g_1(x-\frac{\pi i}{2})g_1(x+\frac{\pi i}{2}) & = 1\\
g_2(x-\frac{\pi i}{2})g_2(x+\frac{\pi i}{2}) &= G_1(x).
\end{align}
\end{subequations}
One can show that the ratio $g_2(x)/g_1(x)$ is free of zeros and
poles for $|\Im(x)|<\pi$.
Similiarly, $G_1(x)$ is analytic and non-zero in $|\Im(x)|<\frac{\pi}{2}$.
Thus solving the functional relation for $g_2(x)$ using Fourier series, we
find
\begin{equation}
\log{g_2(x)} = \log{g_1(x)} + \veps*\log{G_1}(x)
\end{equation}
where the kernel in the convolution is
\begin{equation}
\veps(x) = \frac{\h_2(0,t^{2\nu})\h_3(0,t^{2\nu})\h_3(ix,t^{2\nu})}
{2\pi\h_2(ix,t^{2\nu})}
\end{equation}
and $\nu=5/4$. We do not need the explicit solution $g_2(x)$ since we 
only need to
evaluate it in the scaling limit.

The functional relations for the finite-size corrections $l(u)$ are 
obtained from
\eqref{e:functional} using \eqref{e:f3} and \eqref{e:g3}
\begin{subequations}\label{e:l3}
\begin{align}
l_1(x-\frac{\pi i}{2})l_1(x+\frac{\pi i}{2}) &= T_2(x)\\
l_2(x-\frac{\pi i}{2})l_2(x+\frac{\pi i}{2}) &= \frac{T_1(x)}{G_1(x)}.
\end{align}
\end{subequations}
To solve for $\log l_1(x)$ and $\log l_2(x)$ we need to remove the 
singularities
arising from the zeros in the interior of strips~1 and 2, that is, the
$m_1$ and $m_2$ 1-strings $\{\frac{\l}{2}\pm{iv^{(1)}_j}\}$
and $\{3\l\pm{iv^{(2)}_k}\}$.
Using elementary solutions of the inversion relation
$l(x-\frac{\pi i}{2})l(x+\frac{\pi i}{2})=1$ with a single zero inside the
analyticity strip we find
\begin{equation}
l_i(x)\prod_{j=1}^{m_i}
\frac{
\h_2(\frac{ix}{2}+\frac{5}{2}i v^{(i)}_j,t^{5/4})
\;\h_2(\frac{ix}{2}-\frac{5}{2}i v^{(i)}_j,t^{5/4})}
{\h_1(\frac{ix}{2}+\frac{5}{2}i v^{(i)}_j,t^{5/4})
\;\h_1(\frac{ix}{2}-\frac{5}{2}i v^{(i)}_j,t^{5/4})},
\qquad \text{$i = 1, 2$}
\end{equation}
is free of zeros and poles inside each strip.
Applying Fourier series to the logarithms of \eqref{e:l3} and using Fourier
inversion thus gives the following nonlinear integral equations valid for
$|\Im(x)|<\pi$
\begin{subequations}\label{e:logl3}
\begin{eqnarray}
\log{l_1(x)}\!\!\! &=&\! \!\!\sum_{j=1}^{m_1}\log{\frac{
\h_1(\frac{ix}{2}+\frac{5}{2}i v^{(1)}_j,t^{\frac{5}{4}})
\;\h_1(\frac{ix}{2}-\frac{5}{2}i v^{(1)}_j,t^{\frac{5}{4}})}
{\h_2(\frac{ix}{2}+\frac{5}{2}i v^{(1)}_j,t^{\frac{5}{4}})
\;\h_2(\frac{ix}{2}-\frac{5}{2}i v^{(1)}_j,t^{\frac{5}{4}})}}
+ {\veps}*\log{T_2}(x)+C_1\\
\log{l_2(x)}\!\!\! &=&\!\!\!
\sum_{k=1}^{m_2}\log{\frac{
\h_1(\frac{ix}{2}+\frac{5}{2}i v^{(2)}_k,t^{\frac{5}{4}})
\;\h_1(\frac{ix}{2}-\frac{5}{2}i v^{(2)}_k,t^{\frac{5}{4}})}
{\h_2(\frac{ix}{2}+\frac{5}{2}i v^{(2)}_k,t^{\frac{5}{4}})
\;\h_2(\frac{ix}{2}-\frac{5}{2}i v^{(2)}_k,t^{\frac{5}{4}})}}
+ {\veps}*\log{\frac{T_1(x)}{G_1(x)}}+C_2
\label{TBA}
\end{eqnarray}
\end{subequations}
where the integration constants $C_1$ and $C_2$ are multiples of $\pi 
i$ related to the
choices of branches for the logarithms. By going to the critical 
limit $t\to 0$, and
comparing these equations with the corresponding equations in 
\cite{OPW97}, one finds that
$C_1=C_2=0$. In effect we are fixing the branches of the logarithms 
of $l_j$ exactly as in
the critical case.

We next write these equations in terms of $t_1(x)$ and $t_2(x)$ as
\begin{equation}\label{e:logt3}
\begin{split}
\log{t_1(x)}& = \log{f_1(x)} + \log{g_1(x)} + {\veps}*\log{T_2}(x)\\
&\qquad +\sum_{j=1}^{m_1}\log{\frac{
\h_1(\frac{ix}{2}+\frac{5}{2}i v^{(1)}_j,t^{\frac{5}{4}})
\;\h_1(\frac{ix}{2}-\frac{5}{2}i v^{(1)}_j,t^{\frac{5}{4}})}
{\h_2(\frac{ix}{2}+\frac{5}{2}i v^{(1)}_j,t^{\frac{5}{4}})
\;\h_2(\frac{ix}{2}-\frac{5}{2}i v^{(1)}_j,t^{\frac{5}{4}})}}\\
\log{t_2(x)}& = \log{f_2(x)} + \log{g_2(x)} + {\veps}*\log{T_1}(x)
-{\veps}*\log{G_1}(x)+\\
&\qquad + \sum_{k=1}^{m_2}\log{\frac{
\h_1(\frac{ix}{2}+\frac{5}{2}i v^{(2)}_k,t^{\frac{5}{4}})
\;\h_1(\frac{ix}{2}-\frac{5}{2}i v^{(2)}_k,t^{\frac{5}{4}})}
{\h_2(\frac{ix}{2}+\frac{5}{2}i v^{(2)}_k,t^{\frac{5}{4}})
\;\h_2(\frac{ix}{2}-\frac{5}{2}i v^{(2)}_k,t^{\frac{5}{4}})}}.\\
\end{split}
\end{equation}

We are interested in the solutions of these equations
in the scaling limit. Replacing $t^{\frac{5}{4}}$ by $\frac{\mu}{N}$ 
we see that all
dependence on $t$ disappears and only a dependence on $N$ remains. We 
assume the
relevant functions have the general scaling form
\begin{equation}
\hat{h}(x)=\lim_{N\to\infty}h(x+\ln N)
\end{equation}
and set
\begin{equation}\label{e:e}
e^{-\u_i(\h)} = \lim_{N\rightarrow\infty}t_i(\h-\log{\frac{\mu}{N}})
  = \hat{t}_j(\h-\log{\mu}),\qquad i = 1,2
\end{equation}
with $\mu=\frac{mR}{4}>0$. The $\u_i(\h)$ are precisely the 
pseudo-energies and $\h$ is the
rapidity.

Now taking the
scaling limit of the nonlinear integral equations using \eqref{e:f3}, 
\eqref{e:g3}
\eqref{e:logl3} and \eqref{e:e} gives the excited TBA equations
\begin{equation}
\label{e:TBA}
\begin{split}
\u_1(\h)& = -\log{\tanh^2{\frac{\h}{2}}}
-\sum_{j=1}^{m_1}\log\Bigl[\tanh\big(\frac{\h}{2}+\frac{\b^{(1)}_j}{2}\big)
\tanh\big(\frac{\h}{2}-\frac{\b^{(1)}_j}{2}\big)\Bigr]\\
&- \frac{1}{2\pi}\int_{-\infty}^{\infty}\!d\h'\,\frac{\log(1+e^{-\u_2(\h')})}
{\cosh(\h-\h')}\\
\u_2(\h)& = 2mR\cosh\h-\log{\tanh^2{\frac{\h}{2}}}
-\sum_{k=1}^{m_2}\log\Bigl[\tanh\big(\frac{\h}{2}+\frac{\b^{(2)}_k}{2}\big)
\tanh\big(\frac{\h}{2}-\frac{\b^{(2)}_k}{2}\big)\Bigr]\\
&- \frac{1}{2\pi}\int_{-\infty}^{\infty}\!d\h'\,\frac{\log(1+e^{-\u_1(\h')})}
{\cosh(\h-\h')}
\end{split}
\end{equation}
where we have moved to the scaled locations of zeros $\beta_j^{(i)}$ 
as given by (\ref{yz}). 

The excited TBA equations contain $m_1+m_2$ extra parameters which 
are the locations
of the zeros inside strips~1 and 2.  These can be determined by considering the
scaling limit of the TBA functional equations
\begin{align}\label{e:t3}
t_1(x-i\frac{\pi}{2})t_1(x+i\frac{\pi}{2})& = 1 + t_2(x)\\
t_2(x-i\frac{\pi}{2})t_2(x+i\frac{\pi}{2})& = 1 + t_1(x).
\end{align}
Setting $x=\frac{\pi i}{2}+5v_j^{(i)}$ we see that the LHS must vanish. Hence
one can show that in the scaling limit
\begin{align}\label{e:-1}
&\hat{t}_2(\b^{(1)}_j-\frac{\pi i}{2}-\log\mu) =
-1 = e^{n_j^{(1)}\pi{i}},\qquad j=1,2,\ldots,m_1\\
&\hat{t}_1(\b^{(2)}_k-\frac{\pi i}{2}-\log\mu) =
-1 = e^{n_k^{(2)}\pi{i}},\qquad k=1,2,\ldots,m_2
\end{align}
where $n_j^{(1)}$ and $n_k^{(2)}$ are odd integers.
Moreover these integers, which are determined by windings, must be 
precisely the same
as in the critical case $(R=0)$, namely,
\begin{align}
n_j^{(1)}& =
2(m_1-j)-m_2 + 1 + 2I_j^{(1)},\qquad j=1,2,\ldots,m_1\\
n_k^{(2)}& =
2(m_2-k)-m_1 + 1 + 2I_k^{(2)},\qquad k=1,2,\ldots,m_2.
\end{align}
Applying \eqref{e:e}, the auxiliary conditions determining the 
locations of zeros
become
\begin{align}\label{e:AUX1}
\u_2(\b^{(1)}_j-\frac{\pi i}{2}) &= n_j^{(1)}\pi{i},\qquad j=1,2,\ldots,m_1\\
\u_1(\b^{(2)}_k-\frac{\pi i}{2}) &= n_k^{(2)}\pi{i},\qquad k=1,2,\ldots,m_2.
\end{align}

For numerical purposes we need a more explicit form of the auxiliary equations
obtained by replacing $\h$ with $\b_j^{(i)}-\frac{\pi i}{2}$ in the 
TBA equations
\begin{equation}
\begin{split}\label{aux}
&-2mR\sinh\beta^{(1)}_{j}=\int{d\h\over{2\pi}}\,
{\log\left(1+e^{-\epsilon_1}(\h)\right)\over{\sinh(\beta_{j}^{(1)}
-\h)}}-i\sum_{k=1}^{m_2}\log\Big[\tanh\Big({\pi i\over{4}}
+{\beta^{(2)}_{k}-\beta_{j}^{(1)}\over{2}}\Big)\Big]\\
&\quad\mbox{}-i\sum_{k=1}^{m_2}\log\Big[\tanh\Big({\pi i\over{4}}
-{\beta_{k}^{(2)}+\beta^{(1)}_{j}\over{2}}\Big)\Big]
-i\log\Big[\tanh^2\Big({\pi i\over{4}}-{\beta_{j}^{(1)}\over{2}}
\Big)\Big] +n_{j}^{(1)}\pi\\
&\qquad\qquad\qquad\;\;0=\int{d\h\over{2\pi}}\,
{\log\left(1+e^{-\epsilon_2}(\h)\right)
\over{\sinh(\beta_{k}^{(2)}-\h)}}-i\sum_{j=1}^{m_1}
\log\Big[\tanh\Big({\pi i\over{4}}
+{\beta^{(1)}_{j}-\beta_{k}^{(2)}\over{2}}\Big)\Big]\\
&\quad\mbox{}-i\sum_{j=1}^{m_1}\log\Big[\tanh\Big({\pi i\over{4}}
-{\beta_{k}^{(2)}+\beta^{(1)}_{j}\over{2}}\Big)\Big]
-i\log\Big[\tanh^2\Big({\pi i\over{4}}-{\beta_{k}^{(2)}\over{2}}
\Big)\Big] +n_{k}^{(2)}\pi.
\end{split}
\end{equation}
We propose \eqref{e:TBA}, together with the auxiliary equations
\eqref{aux}, as the TBA equations for all excitations in
the massive perturbation of the tricritical Ising model on a cylinder
with the $(1,1)$ boundary condition.

It remains to relate the finite-size corrections to the pseudo-energies to obtain
the scaled energies.  Again following \cite{OPW97}, one can determine 
the finite-size
corrections to the eigenvalues of the double row transfer matrix from
\eqref{e:logl3} as
\begin{equation}\label{e:finite}
\begin{split}
-\frac{1}{2}\log{D_1(x)}& =
\frac{mR\cosh{x}}{N}\Bigl[\sum_{j=1}^{m_1}2\cosh{\b^{(1)}_j}
-\frac{1}{2\pi}\int_{-\infty}^{\infty}\!d\h
\cosh{\h}\log{(1+e^{-\u_2(\h)})}\Bigr]
\end{split}
\end{equation}
where we have neglected terms of order $o(1/N)$. The scaling energies 
of excitations
are therefore
\begin{equation}\label{xenergies}
\begin{split}
R E(R)& =
2mR\sum_{j=1}^{m_1}\cosh{\b^{(1)}_j}
-\frac{mR}{2\pi}\int_{-\infty}^{\infty}\!d\h\,
\cosh{\h}\log{(1+e^{-\u_2(\h)})}.
\end{split}
\end{equation}

\subsection{Analysis of UV and IR limits}

One can check that the UV limit $R\to 0$ of the massive TBA 
equations reproduces the
``conformal TBA'' equations  of O'Brien, Pearce and Warnaar~\cite{OPW97}. To do
this, one should make the identifications
\begin{equation}\label{rescale}
\vartheta\sim\log{mR\over{4}}+x,\qquad
\beta^{(i)}_{j}\sim\log{mR\over{4}}+y^{(i)}_{j}
\end{equation}
and for the pseudo-energies
\begin{equation}
{\hat\epsilon}(x)\sim\epsilon\left(\log{mR\over{4}}+x\right).
\end{equation}
Doing this we find that this limit indeed reproduces the known 
``conformal TBA''
equations.

The IR limit $R\to\infty$ of the massive TBA equations
can also give many insights on the field theoretic behaviours.
In this limit one can interpret the auxiliary equations \eqref{aux} as
Bethe ansatz equations for $m_1$ massive particles and $m_2$ massless 
particles interacting with each other.
The momenta and energies of the particles determined by the equations 
will introduce corrections in the total energy corresponding to vacuum polarization 
due to a large but finite value of $mR$.

In this limit we find that the TBA equations become
\begin{eqnarray}
\u_2(\h)&\sim& 2mR\cosh\h+O(1)\\
\u_1(\h)&=&\log\Big[\frac{\cosh\h+1}{\cosh\h-1}\Big]+
\sum_{j=1}^{m_1}\log\Big[
\frac{\cosh\h+\cosh\b^{(1)}_j}{\cosh\h-\cosh\b^{(1)}_j}\Big]
\end{eqnarray}
and the first auxiliary equation is simplified as
\begin{equation}
2mR\sinh\b^{(1)}_j=-(m_2+1+n_j^{(1)})\pi+O\big(\frac{1}{R}\big).
\end{equation}
This implies
\begin{equation}\label{b1ir}
\b^{(1)}_j\sim -\frac{(m_2+1+n_j^{(1)})\pi}{2mR}+O\big(\frac{1}{R^2}\big).
\end{equation}
Substituting these results into the second auxiliary equation in \eqref{aux}
we find
\begin{equation}
\u_1(\b^{(2)}_k-\frac{\pi i}{2})=
(m_1+1)\log\Big[\frac{-i\sinh\b^{(2)}_k+1}{-i\sinh\b^{(2)}_k-1}\Big]
+O\big(\frac{1}{R}\big)=n_k^{(2)}\pi i.
\end{equation}
Solving this in the IR limit gives the limiting locations of zeros in strip~2
\begin{equation}\label{b2ir}
\sinh\b^{(2)}_k=\cot\Big(\frac{n_k^{(2)}\pi}{2(m_1+1)}\Big),
\qquad k=1,2,\ldots,m_2.
\end{equation}
Finally, the large $R$ limit of the scaling energy is given by
\begin{equation}
E(R)\sim2m_1 m-\frac{m}{2\pi}\int_{-\infty}^{\infty}d\h\cosh\h\,{\cal S}(\h)
e^{-2mR\cosh\h}
\end{equation}
where ${\cal S}(\h)$ is given by $O(1)$ term in $\u_2(\h)$
\begin{equation}
{\cal S}(\h)=\tanh^{2}\frac{\h}{2}\;\prod_{k=1}^{m_2}\Big[
\frac{\cosh\h-\cosh\b^{(2)}_k}{\cosh\h+\cosh\b^{(2)}_k}\Big]
\exp\int_{-\infty}^{\infty}\frac{d\h'}{2\pi}\;
\frac{\log\big(1+\tanh^{2(m_1+1)}\frac{\h'}{2}\big)}
{\cosh(\h-\h')}.
\end{equation}
The leading term is the energy of $m_1$ massive particles with zero 
momentum and
the second term describes the finite-size vacuum polarization in the 
presence of
$m_2$ massless particles with momenta given by \eqref{b2ir}.
Notice that the contribution from three particle interactions
in the Yang-Lee model \cite{Dorey,BLZ} is absent here 
since the relativistic kink particles do not have bound states.  
The quantum number $m_1$ giving the number of massive particles can be identified as the number of
domain walls or kinks in the configurations  of the RSOS off-critical $A_4$ lattice model in
Regime~III. This identification is possible because the same classification of eigenvalues in terms
of $(\vm,\vn)$ systems applies to the lattice model throughout the off-critical Regime~III. Indeed,
an $(\vm,\vn)$ system appears in (2.5b) and (2.6) of \cite{BaxP83} with $t=m_1$, $s=m_2$, $r=n_1$ and
$p=n_2$. Although the periodic case was considered in this previous paper the same $(\vm,\vn)$ system
applies for the $(r,s)=(1,1)$ boundary condition with double row transfer matrices. The
identification of the number of massive particles $m_1$ with the number of domain walls can therefore
be easily made by looking at low temperature expansions.

\section{Massive Numerics}
Away from the UV and IR limits, 
the excited TBA equations and associated auxiliary equations can only be
solved numerically.  While simple iteration of pseudo-energies is usually enough for
ground-state TBA analysis, extra complications arise for excited TBA equations 
due to the presence of the zeros. 

Our numerical algorithm is to iteratively update the pseudo-energies with previously
determined values of the zeros and then to find new values for the zeros using the
updated  pseudo-energies. This iteration continues until we obtain the required data with
a desired  accuracy. One delicate point arises when one solves the auxiliary
equations.   There is no natural way to rearrange the second equation in \eqref{aux} to
express the zeros in strip 2 directly in terms of other quantities. Instead, we use
the log term $\Phi^{(k)}=-i\log\big[\tanh^2\big({\pi i\over{4}}-{\beta_{k}^{(2)}\over{2}}
\big)\big]$ as iteration variable which gives, in turn by inversion, the improved values
of the zeros $\beta_{k}^{(2)}$ in strip 2. This $\Phi^{(k)}$ is naturally interpreted as
a phase factor. We coded the algorithm in
the ${\rm MATLAB}^{\rm TM}$/Octave programming language. Typical running time on a 500
MHz computer to achieve an accuracy of five decimal digits is about one minute for a
given  value of
$R$.   

For the purposes of plotting numerical data it is convenient to normalize the scaling
energies $RE(R)$. As we have discussed, the leading term in $RE(R)$
diverges linearly with $R$ as $R\to\infty$ while it approaches a constant as $R\to 0$. 
To plot the whole flow from UV to IR in one plot, we use a normalized
scaling function
\begin{equation}\label{newscale}
{\cal E}(R)=\frac{RE(R)}{2(\pi+mR)}
\end{equation} 
with the UV  and IR limits
\begin{equation}
{\cal E}(0)=\lim_{R\to 0}{\cal E}(R)=-\frac{7}{240}+\Delta_{r,s}+n,\qquad
{\cal E}(\infty)=\lim_{R\to\infty}{\cal E}(R)=m_1.
\end{equation} 

\begin{table}[htbp]
\leftline{
\begin{tabular}{|c|c|c|c|c|c||c|c|c|c|c|c|}\hline\hline
\rule[-.2cm]{0cm}{.65cm}$\ m_1\ $&$\ m_2\ $&
$\sum I$&$\ \ \#\ \ $&$\Delta+n$
&${\cal E}(\infty)$
&$\ m_1\ $&$\ m_2\ $&
$\sum I$&$\ \ \#\ \ $&$\Delta+n$
&${\cal E}(\infty)$ \\ \hline\hline
\rule[-.2cm]{0cm}{.65cm}0&0&0&$[1|1]$&0&0
&2&0&9&$[3|5]$&11&2\\ \hline
\rule[-.2cm]{0cm}{.65cm}2&0&0&$[1|1]$&2&2
&4&0&3&$[1|3]$&11&4\\ \hline
\rule[-.2cm]{0cm}{.65cm}2&0&1&$[1|1]$&3&2
&4&2&5&$[1|6]$&11&4\\ \hline
\rule[-.2cm]{0cm}{.65cm}2&0&2&$[2|2]$&4&2
&2&0&10&$[1|6]$&12&2\\ \hline
\rule[-.2cm]{0cm}{.65cm}2&0&3&$[2|2]$&5&2
&4&0&4&$[1|5]$&12&4\\ \hline
\rule[-.2cm]{0cm}{.65cm}2&0&4&$[3|3]$&6&2
&4&2&6&$[1|9]$&12&4\\ \hline
\rule[-.2cm]{0cm}{.65cm}4&2&0&$[1|1]$&6&4
&2&0&11&$[1|6]$&13&2\\ \hline
\rule[-.2cm]{0cm}{.65cm}2&0&5&$[3|3]$&7&2
&4&0&5&$[1|6]$&13&4\\ \hline
\rule[-.2cm]{0cm}{.65cm}4&2&1&$[1|1]$&7&4
&4&2&7&$[1|11]$&13&4\\ \hline
\rule[-.2cm]{0cm}{.65cm}2&0&6&$[4|4]$&8&2
&2&0&12&$[1|7]$&14&2\\ \hline
\rule[-.2cm]{0cm}{.65cm}4&0&0&$[1|1]$&8&4
&4&0&6&$[1|9]$&14&4\\ \hline
\rule[-.2cm]{0cm}{.65cm}4&2&2&$[2|2]$&8&4
&4&2&8&$[1|15]$&14&4\\ \hline
\rule[-.2cm]{0cm}{.65cm}2&0&7&$[4|4]$&9&2
&6&2&0&$[1|1]$&14&6\\ \hline
\rule[-.2cm]{0cm}{.65cm}4&0&1&$[1|1]$&9&4
&2&0&13&$[1|7]$&15&2\\ \hline
\rule[-.2cm]{0cm}{.65cm}4&2&3&$[3|3]$&9&4
&4&0&7&$[1|11]$&15&4\\ \hline
\rule[-.2cm]{0cm}{.65cm}2&0&8&$[3|5]$&10&2
&4&2&9&$[1|17]$&15&4\\ \hline
\rule[-.2cm]{0cm}{.65cm}4&0&2&$[2|2]$&10&4
&6&2&1&$[1|2]$&15&6\\ \hline
\rule[-.2cm]{0cm}{.65cm}4&2&4&$[1|5]$&10&4
& & & & & &\\ \hline
\end{tabular}
}
\caption{Energy levels for the $(r,s)=(1,1)$ boundary condition with $\Delta=0$. The
quantum numbers $m_1$, $m_2$, $\sum I=\sum I^{(1)}+\sum I^{(2)}$ are shown along with the
conformal-massive dictionary connecting the UV conformal data $\Delta+n$ with the number of massive
particles ${\cal E}(\infty)=m_1$ in the IR limit. The degeneracies 
$\#=[\ell_1|\ell_2]$ indicate that $\ell_1$ levels are plotted out of the $\ell_2$ levels with the
given quantum numbers. The plotted energy levels are complete in the conformal
limit up to $n=9$. 
} 
\end{table}

\begin{figure}[p]
\begin{center}
\includegraphics[width=.9\linewidth]{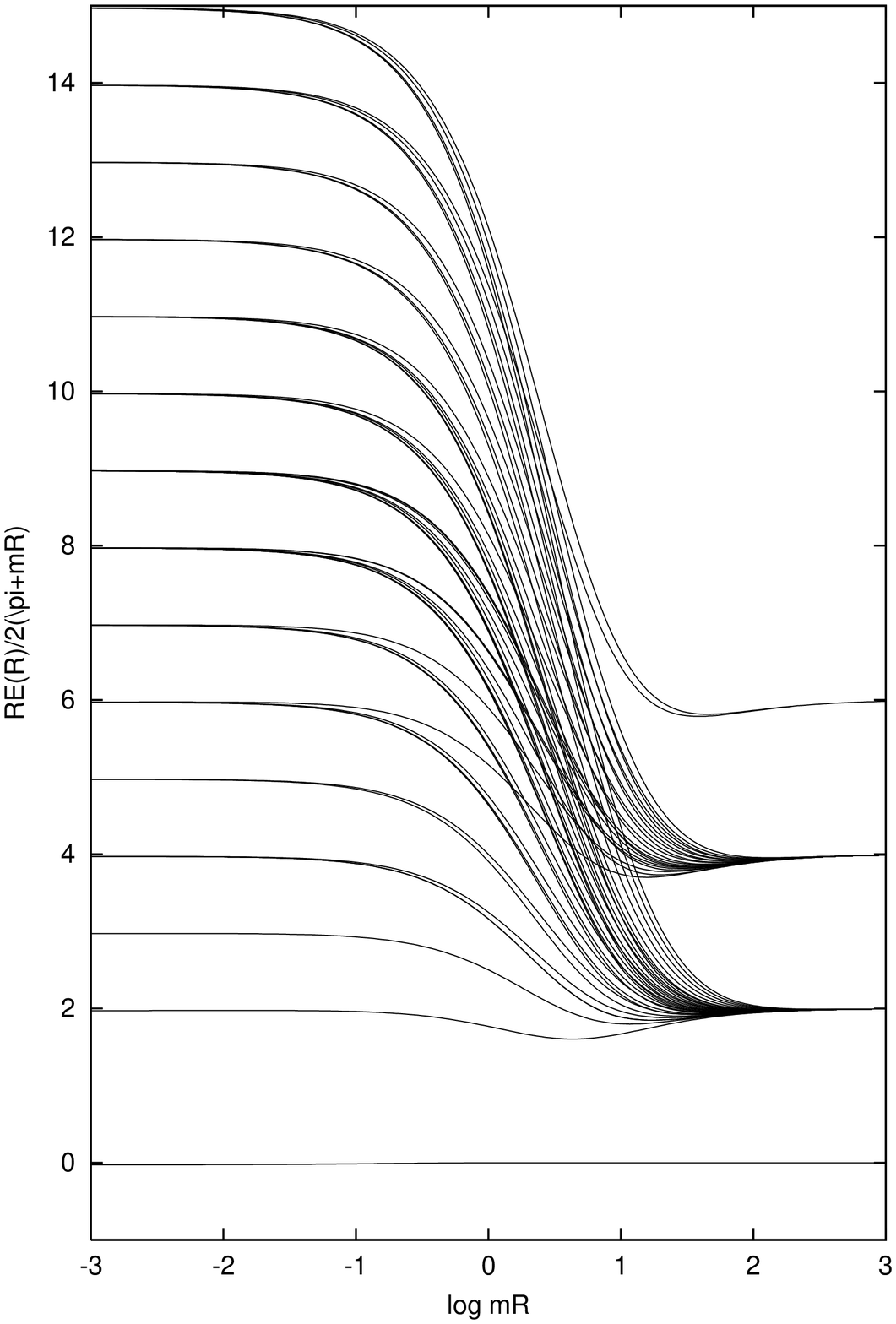}
\end{center}
\caption{Normalized scaling energies ${\cal E}(R)=\frac{RE(R)}{2(\pi+mR)}$ plotted against $\log_{10}
mR$ for the
$(r,s)=(1,1)$ boundary condition.}
\label{r1s1}
\end{figure}
\begin{figure}[p]
\begin{center}
\vspace{-.3truein}
\includegraphics[width=.9\linewidth]{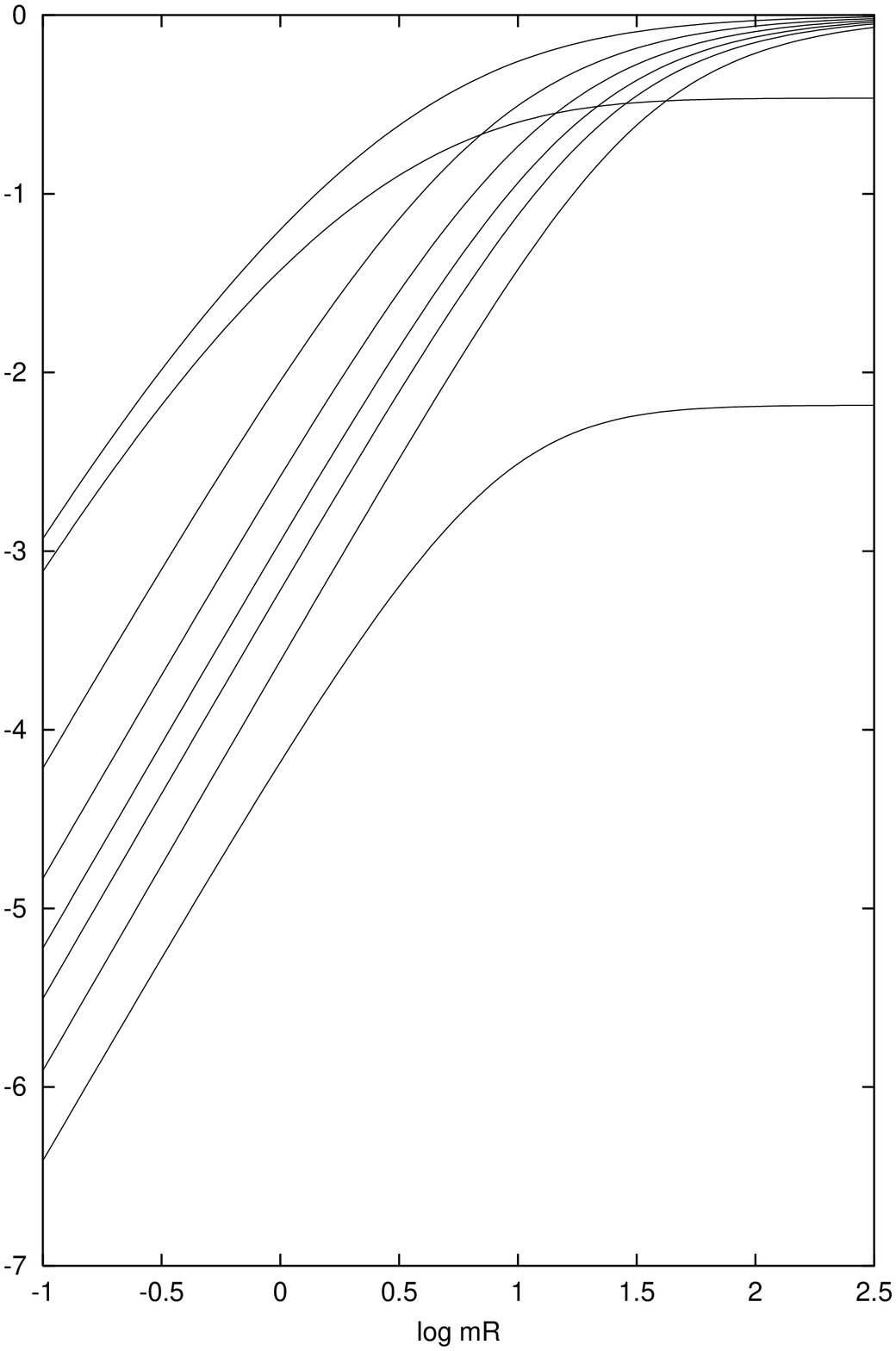}
\end{center}
\vspace{-.1in}
\caption{Locations of zeros plotted against $\log_{10} mR$ for a typical energy level for the
$(r,s)=(1,1)$ boundary condition. Here $m_1=6$, $m_2=2$ and 
$(I^{(1)}|I^{(2)})=(1,0,0,0,0,0|1,0)$. As $R\to\infty$, the six zeros in strip~1 approach zero
while the two zeros in strip~2 converge to constants given by (\ref{b2ir}). Since the zeros
are in different strips there are no collisions of zeros.}
\label{zeroplot}
\end{figure}

\subsection{Sector $(r,s)=(1,1)$}
In Figure~5 we show our numerical results for the $(r,s)=(1,1)$ boundary condition.
This sector has an even number of zeros in each of the two strips.
The vertical axis is the normalized scaling function ${\cal E}(R)$ and the 
horizontal axis is $\log_{10}(mR)$. 
We plot selected normalized scaling energies for up to $m_1=6$ zeros in strip 1 and $m_2=2$
zeros in strip 2 for allowed quantum numbers including the lowest 30
levels. Note that eigenvalues which are degenerate in the conformal UV limit become non-degenerate
when the perturbing thermal field is turned on and the conformal symmetry is broken. Table~1
summarizes how the UV descendant levels flow into the IR particle states.  The plotted energy levels
are complete  in the conformal limit up to $n=9$ corresponding to the expansion of the finitized
Virasoro character (\ref{FinVirChar}) in the limit $N\to\infty$
\begin{eqnarray}
q^{c/24}\chi_{1,1}(q)\!\!&=&\!\!1+q^2\qbin{\infty}{2}\qbin{1}{0}+q^8\qbin{\infty}{4}\qbin{2}{0}
+q^6\qbin{\infty}{4}\qbin{2}{2}
+q^{18}\qbin{\infty}{6}\qbin{3}{0}+q^{14}\qbin{\infty}{6}\qbin{3}{2}+\ldots\nonumber\\
&=&\!\!1+q^2(1+q+2q^2+2q^3+3q^4+3q^5+4q^6+4q^7+\ldots)\nonumber\\
&&\quad\mbox{}\!\!+q^8(1+q+\ldots)+q^6(1+q+2q^2+3q^3+\ldots)+\ldots\\
&=&\!\!1+q^2+q^3+2q^4+2q^5+4q^6+4q^7+7q^8+8q^9+\ldots\nonumber
\end{eqnarray}

The location of zeros also flow as $R$ changes. To illustrate this, we plot in
Figure~6 the locations of the eight zeros for the energy level with $m_1=6$, $m_2=2$ and
quantum numbers
$(I^{(1)}|I^{(2)})=(1,0,0,0,0,0|1,0)$.
Near the UV limit, the zeros are linear in $\log(mR)$ as expected in 
\eqref{rescale}. One can check that the constant separations between
them agree with those from the ``conformal TBA'' equations.
The IR behaviour of the zeros is also as expected. 
The six zeros in strip~1 decay exponentially to 0 as plotted against $\log(mR)$. 
This is in accord with the fact that the zeros approach 0 as $1/mR$ and the 
constants of proportionality are confirmed to be precisely those in \eqref{b1ir}. The
massive particle states become frozen in the IR limit and accordingly the two zeros in
strip~2 converge exponentially to the finite constant values given by \eqref{b2ir}. The
massless particles have prescribed momenta in the same limit.
These behaviours are reproduced consistently in the analysis of all other
boundary conditions and quantum numbers.

\subsection{Sector $(r,s)=(3,1)$}

In this sector, the $(\vm,\vn)$ system is
\begin{equation}
\vm + \vn = \frac{1}{2}\big[(N+1)\ve_{1}+\vI\vm\big]
\end{equation}
and $N$ and $m_1$ are even while $m_2$ is odd. Hence there is an even number of zeros in strip~1 and
an odd number in strip~2. Repeating the derivation of the TBA equations leads to the same equations
as for the $(r,s)=(1,1)$ boundary condition but with $e^{-\u_2(\h)}$ replaced with $-e^{-\u_2(\h)}$.
The TBA equations in this sector thus become
\begin{equation}
\label{e:TBA31}
\begin{split}
\u_1(\h)& = -\log{\tanh^2{\frac{\h}{2}}}
-\sum_{j=1}^{m_1}\log\Bigl[\tanh\big(\frac{\h}{2}+\frac{\b^{(1)}_j}{2}\big)
\tanh\big(\frac{\h}{2}-\frac{\b^{(1)}_j}{2}\big)\Bigr]\\
&- \frac{1}{2\pi}\int_{-\infty}^{\infty}\!d\h'\,\frac{\log(1-e^{-\u_2(\h')})}
{\cosh(\h-\h')}\\
\u_2(\h)& = 2mR\cosh\h-\log{\tanh^2{\frac{\h}{2}}}
-\sum_{k=1}^{m_2}\log\Bigl[\tanh\big(\frac{\h}{2}+\frac{\b^{(2)}_k}{2}\big)
\tanh\big(\frac{\h}{2}-\frac{\b^{(2)}_k}{2}\big)\Bigr]\\
&- \frac{1}{2\pi}\int_{-\infty}^{\infty}\!d\h'\,\frac{\log(1+e^{-\u_1(\h')})}
{\cosh(\h-\h')}.
\end{split}
\end{equation}
The auxiliary equations are 
\begin{equation}
\begin{split}\label{aux31}
&-2mR\sinh\beta^{(1)}_{j}=\int{d\h\over{2\pi}}\,
{\log\left(1+e^{-\epsilon_1}(\h)\right)\over{\sinh(\beta_{j}^{(1)}
-\h)}}-i\sum_{k=1}^{m_2}\log\Big[\tanh\Big({\pi i\over{4}}
+{\beta^{(2)}_{k}-\beta_{j}^{(1)}\over{2}}\Big)\Big]\\
&\quad\mbox{}-i\sum_{k=1}^{m_2}\log\Big[\tanh\Big({\pi i\over{4}}
-{\beta_{k}^{(2)}+\beta^{(1)}_{j}\over{2}}\Big)\Big]
-i\log\Big[\tanh^2\Big({\pi i\over{4}}-{\beta_{j}^{(1)}\over{2}}
\Big)\Big] +n_{j}^{(1)}\pi\\
&\qquad\qquad\qquad\;\;0=\int{d\h\over{2\pi}}\,
{\log\left(1-e^{-\epsilon_2}(\h)\right)
\over{\sinh(\beta_{k}^{(2)}-\h)}}-i\sum_{j=1}^{m_1}
\log\Big[\tanh\Big({\pi i\over{4}}
+{\beta^{(1)}_{j}-\beta_{k}^{(2)}\over{2}}\Big)\Big]\\
&\quad\mbox{}-i\sum_{j=1}^{m_1}\log\Big[\tanh\Big({\pi i\over{4}}
-{\beta_{k}^{(2)}+\beta^{(1)}_{j}\over{2}}\Big)\Big]
-i\log\Big[\tanh^2\Big({\pi i\over{4}}-{\beta_{k}^{(2)}\over{2}}
\Big)\Big] +n_{k}^{(2)}\pi
\end{split}
\end{equation}
and the scaling energy becomes
\begin{equation}\label{xenergies31}
\begin{split}
R E(R)& =
2mR\sum_{j=1}^{m_1}\cosh{\b^{(1)}_j}
-\frac{mR}{2\pi}\int_{-\infty}^{\infty}\!d\h\,
\cosh{\h}\log{(1-e^{-\u_2(\h)})}.
\end{split}
\end{equation}

\begin{table}[p]
\leftline{
\begin{tabular}{|c|c|c|c|c|c||c|c|c|c|c|c|}\hline\hline
\rule[-.2cm]{0cm}{.65cm}$\ m_1\ $&$\ m_2\ $&
$\sum I$&$\ \ \#\ \ $&$\Delta+n$
&${\cal E}(\infty)$
&$\ m_1\ $&$\ m_2\ $&
$\sum I$&$\ \ \#\ \ $&$\Delta+n$
&${\cal E}(\infty)$ \\ \hline\hline
\rule[-.2cm]{0cm}{.65cm}2&1&0&$[1|1]$&$\frac{3}{2}$&2
&4&1&1&$[2|2]$&$\frac{3}{2}+6$&4\\ \hline
\rule[-.2cm]{0cm}{.65cm}2&1&1&$[1|1]$&$\frac{3}{2}+1$&2
&2&1&7&$[4|4]$&$\frac{3}{2}+7$&2\\ \hline
\rule[-.2cm]{0cm}{.65cm}2&1&2&$[2|2]$&$\frac{3}{2}+2$&2
&4&1&2&$[3|3]$&$\frac{3}{2}+7$&4\\ \hline
\rule[-.2cm]{0cm}{.65cm}2&1&3&$[2|2]$&$\frac{3}{2}+3$&2
&2&1&8&$[5|5]$&$\frac{3}{2}+8$&2\\ \hline
\rule[-.2cm]{0cm}{.65cm}2&1&4&$[3|3]$&$\frac{3}{2}+4$&2
&4&1&3&$[5|5]$&$\frac{3}{2}+8$&2\\ \hline
\rule[-.2cm]{0cm}{.65cm}2&1&5&$[3|3]$&$\frac{3}{2}+5$&2
&2&1&9&$[5|5]$&$\frac{3}{2}+9$&2\\ \hline
\rule[-.2cm]{0cm}{.65cm}4&1&0&$[1|1]$&$\frac{3}{2}+5$&4
&4&1&4&$[8|8]$&$\frac{3}{2}+9$&4\\ \hline
\rule[-.2cm]{0cm}{.65cm}2&1&6&$[4|4]$&$\frac{3}{2}+6$&2
& & & & & &\\ \hline
\end{tabular}
}
\caption{Energy levels for the $(r,s)=(3,1)$ boundary condition with $\Delta=3/2$. The
quantum numbers $m_1$, $m_2$, $\sum I=\sum I^{(1)}+\sum I^{(2)}$ are shown along with the
conformal-massive dictionary connecting the UV conformal data $\Delta+n$ with the number of massive
particles ${\cal E}(\infty)=m_1$ in the IR limit. The degeneracies 
$\#=[\ell_1|\ell_2]$ indicate that $\ell_1$ levels are plotted out of the $\ell_2$ levels with the
given quantum numbers. The plotted energy levels are complete in the conformal
limit up to $n=9$.} 
\end{table}
\begin{table}[p]
\leftline{
\begin{tabular}{|c|c|c|c|c|c||c|c|c|c|c|c|}\hline\hline
\rule[-.2cm]{0cm}{.65cm}$\ m_1\ $&$\ m_2\ $&
$\sum I$&$\ \ \#\ \ $&$\Delta+n$
&${\cal E}(\infty)$
&$\ m_1\ $&$\ m_2\ $&
$\sum I$&$\ \ \#\ \ $&$\Delta+n$
&${\cal E}(\infty)$ \\ \hline\hline
\rule[-.2cm]{0cm}{.65cm}1&1&0&$[1|1]$&$\frac{7}{16}$&1
&1&1&7&$[1|1]$&$\frac{7}{16}+7$&1\\ \hline
\rule[-.2cm]{0cm}{.65cm}1&1&1&$[1|1]$&$\frac{7}{16}+1$&1
&3&1&4&$[5|7]$&$\frac{7}{16}+7$&3\\ \hline
\rule[-.2cm]{0cm}{.65cm}1&1&2&$[1|1]$&$\frac{7}{16}+2$&1
&1&1&8&$[1|1]$&$\frac{7}{16}+8$&1\\ \hline
\rule[-.2cm]{0cm}{.65cm}1&1&3&$[1|1]$&$\frac{7}{16}+3$&1
&3&1&5&$[3|9]$&$\frac{7}{16}+8$&3\\ \hline
\rule[-.2cm]{0cm}{.65cm}3&1&0&$[1|1]$&$\frac{7}{16}+3$&3
&1&1&9&$[1|1]$&$\frac{7}{16}+9$&1\\ \hline
\rule[-.2cm]{0cm}{.65cm}1&1&4&$[1|1]$&$\frac{7}{16}+4$&1
&3&1&6&$[9|12]$&$\frac{7}{16}+9$&3\\ \hline
\rule[-.2cm]{0cm}{.65cm}3&1&1&$[2|2]$&$\frac{7}{16}+4$&3
&5&3&0&$[1|1]$&$\frac{7}{16}+9$&5\\ \hline
\rule[-.2cm]{0cm}{.65cm}1&1&5&$[1|1]$&$\frac{7}{16}+5$&1
&1&1&10&$[1|1]$&$\frac{7}{16}+10$&1\\ \hline
\rule[-.2cm]{0cm}{.65cm}3&1&2&$[3|3]$&$\frac{7}{16}+5$&3
&3&1&7&$[6|15]$&$\frac{7}{16}+10$&3\\ \hline
\rule[-.2cm]{0cm}{.65cm}1&1&6&$[1|1]$&$\frac{7}{16}+6$&1
&5&1&0&$[1|1]$&$\frac{7}{16}+10$&5\\ \hline
\rule[-.2cm]{0cm}{.65cm}3&1&3&$[5|5]$&$\frac{7}{16}+6$&3
&5&3&1&$[1|1]$&$\frac{7}{16}+10$&5\\ \hline
\end{tabular}
}
\caption{Energy levels for the $(r,s)=(2,1)$ boundary condition with $\Delta=7/16$. The
quantum numbers $m_1$, $m_2$, $\sum I=\sum I^{(1)}+\sum I^{(2)}$ are shown along with the
conformal-massive dictionary connecting the UV conformal data $\Delta+n$ with the number of massive
particles ${\cal E}(\infty)=m_1$ in the IR limit. The degeneracies 
$\#=[\ell_1|\ell_2]$ indicate that $\ell_1$ levels are plotted out of the $\ell_2$ levels with the
given quantum numbers. The plotted energy levels are complete in the conformal
limit up to $n=6$.} 
\end{table}

\begin{figure}[p]
\begin{center}
\includegraphics[width=.9\linewidth]{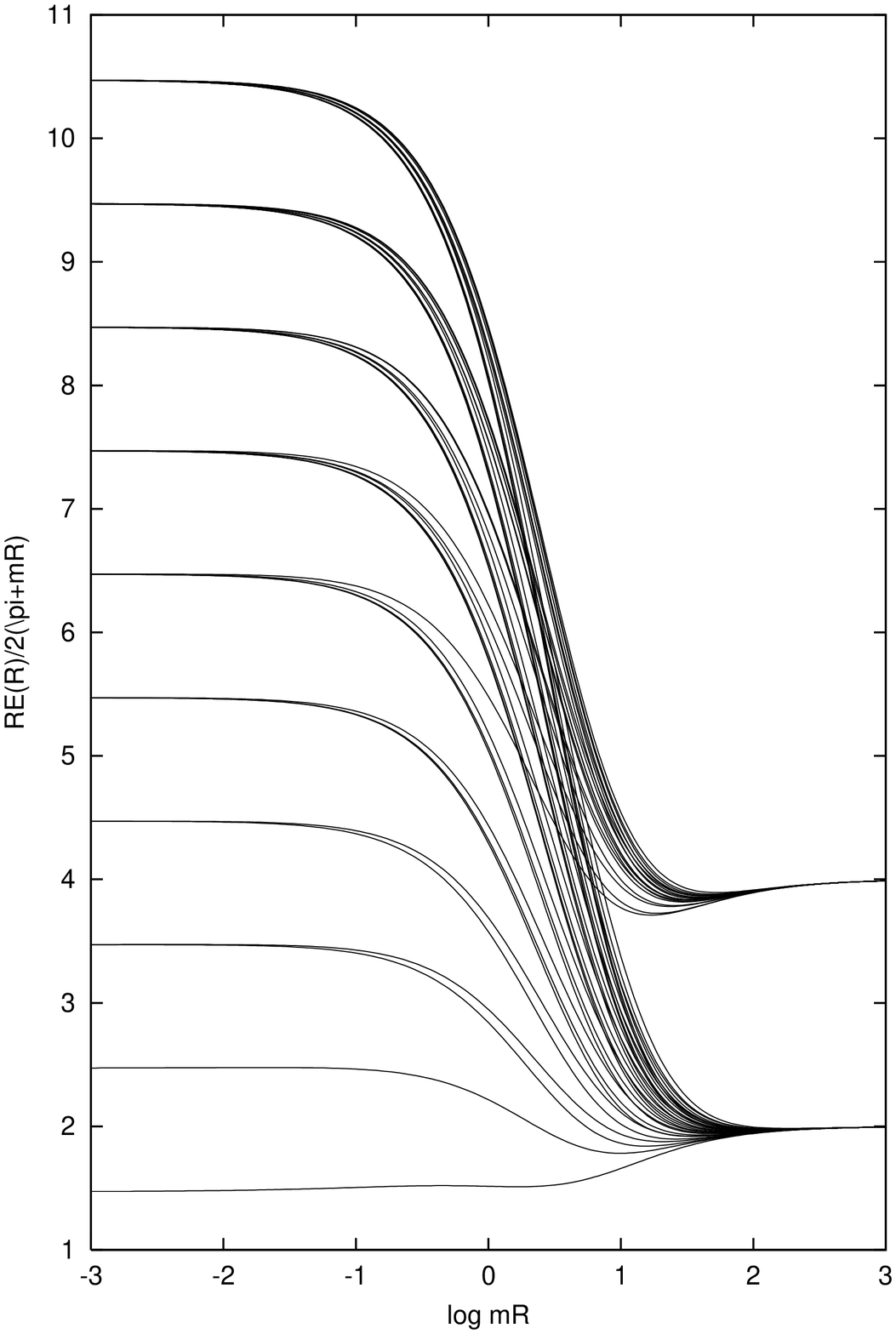}
\end{center}
\caption{Normalized scaling energies ${\cal E}(R)=\frac{RE(R)}{2(\pi+mR)}$ plotted against $\log_{10}
mR$ for the
$(r,s)=(3,1)$ boundary condition.}
\label{r3s1}
\end{figure}

In Figure~7 we show our numerical results for the $(r,s)=(3,1)$ boundary condition.
The vertical axis is the normalized scaling function ${\cal E}(R)$ and the 
horizontal axis is $\log_{10}(mR)$.  We plot selected normalized scaling energies for up to $m_1=4$
zeros in strip~1 and
$m_2=1$ zero in strip~2 for allowed quantum numbers including the lowest 49 
levels.  Table~2 summarizes how the UV descendant levels flow into the IR particle states.  The
plotted energy levels are complete  in the conformal limit up to $n=9$ corresponding to the expansion
of the finitized Virasoro character in the limit $N\to\infty$
\begin{eqnarray}
q^{c/24}\chi_{3,1}(q)&=&q^{3/2}\qbin{\infty}{2}\qbin{1}{1}+q^{3/2+5}\qbin{\infty}{4}\qbin{2}{1}
+\ldots\nonumber\\
&=&q^{3/2}(1+q+2q^2+2q^3+3q^4+3q^5+4q^6+4q^7+5q^8+5q^9\ldots)\nonumber\\
&&\quad\mbox{}+q^{3/2+5}(1+q+2q^2+3q^3+5q^4+\ldots)(1+q)+\ldots\\
&=&q^{3/2}(1+q+2q^2+2q^3+3q^4+4q^5+6q^6+7q^7+10q^8+13q^9\ldots)\nonumber
\end{eqnarray}

\subsection{Sector $(r,s)=(2,1)$}

In this sector, the $(\vm,\vn)$ system is
\begin{equation}
\vm + \vn = \frac{1}{2}(N\ve_{1}+\ve_2+\vI\vm)
\end{equation}
and $N$, $m_1$ and $m_2$ are all odd. Hence there is an odd number of zeros in strip~1 and in strip~2.
Repeating the derivation of the TBA equations leads to essentially the same equations as for the
$(r,s)=(1,1)$ boundary condition but with $e^{-\u_i(\h)}$ replaced with $-e^{-\u_i(\h)}$, $i=1,2$ and
an additional $\pi$ in the second auxiliary equation. The TBA equations in this sector are in fact
\begin{equation}
\label{e:TBA21}
\begin{split}
\u_1(\h)& = -\log{\tanh^2{\frac{\h}{2}}}
-\sum_{j=1}^{m_1}\log\Bigl[\tanh\big(\frac{\h}{2}+\frac{\b^{(1)}_j}{2}\big)
\tanh\big(\frac{\h}{2}-\frac{\b^{(1)}_j}{2}\big)\Bigr]\\
&- \frac{1}{2\pi}\int_{-\infty}^{\infty}\!d\h'\,\frac{\log(1-e^{-\u_2(\h')})}
{\cosh(\h-\h')}\\
\u_2(\h)& = 2mR\cosh\h-\log{\tanh^2{\frac{\h}{2}}}
-\sum_{k=1}^{m_2}\log\Bigl[\tanh\big(\frac{\h}{2}+\frac{\b^{(2)}_k}{2}\big)
\tanh\big(\frac{\h}{2}-\frac{\b^{(2)}_k}{2}\big)\Bigr]\\
&- \frac{1}{2\pi}\int_{-\infty}^{\infty}\!d\h'\,\frac{\log(1-e^{-\u_1(\h')})}
{\cosh(\h-\h')}
\end{split}
\end{equation}
and the auxiliary equations
\begin{equation}
\begin{split}\label{aux21}
&-2mR\sinh\beta^{(1)}_{j}=\int{d\h\over{2\pi}}\,
{\log\left(1-e^{-\epsilon_1}(\h)\right)\over{\sinh(\beta_{j}^{(1)}
-\h)}}-i\sum_{k=1}^{m_2}\log\Big[\tanh\Big({\pi i\over{4}}
+{\beta^{(2)}_{k}-\beta_{j}^{(1)}\over{2}}\Big)\Big]\\
&\quad\mbox{}-i\sum_{k=1}^{m_2}\log\Big[\tanh\Big({\pi i\over{4}}
-{\beta_{k}^{(2)}+\beta^{(1)}_{j}\over{2}}\Big)\Big]
-i\log\Big[\tanh^2\Big({\pi i\over{4}}-{\beta_{j}^{(1)}\over{2}}
\Big)\Big] +n_{j}^{(1)}\pi\\
&\qquad\qquad\qquad\;\;0=\int{d\h\over{2\pi}}\,
{\log\left(1-e^{-\epsilon_2}(\h)\right)
\over{\sinh(\beta_{k}^{(2)}-\h)}}-i\sum_{j=1}^{m_1}
\log\Big[\tanh\Big({\pi i\over{4}}
+{\beta^{(1)}_{j}-\beta_{k}^{(2)}\over{2}}\Big)\Big]\\
&\quad\mbox{}-i\sum_{j=1}^{m_1}\log\Big[\tanh\Big({\pi i\over{4}}
-{\beta_{k}^{(2)}+\beta^{(1)}_{j}\over{2}}\Big)\Big]
-i\log\Big[\tanh^2\Big({\pi i\over{4}}-{\beta_{k}^{(2)}\over{2}}
\Big)\Big] +(n_{k}^{(2)}+1)\pi
\end{split}
\end{equation}
with the scaling energy
\begin{equation}\label{xenergies21}
\begin{split}
R E(R)& =
2mR\sum_{j=1}^{m_1}\cosh{\b^{(1)}_j}
-\frac{mR}{2\pi}\int_{-\infty}^{\infty}\!d\h\,
\cosh{\h}\log{(1-e^{-\u_2(\h)})}.
\end{split}
\end{equation}

\begin{figure}[p]
\begin{center}
\includegraphics[width=.9\linewidth]{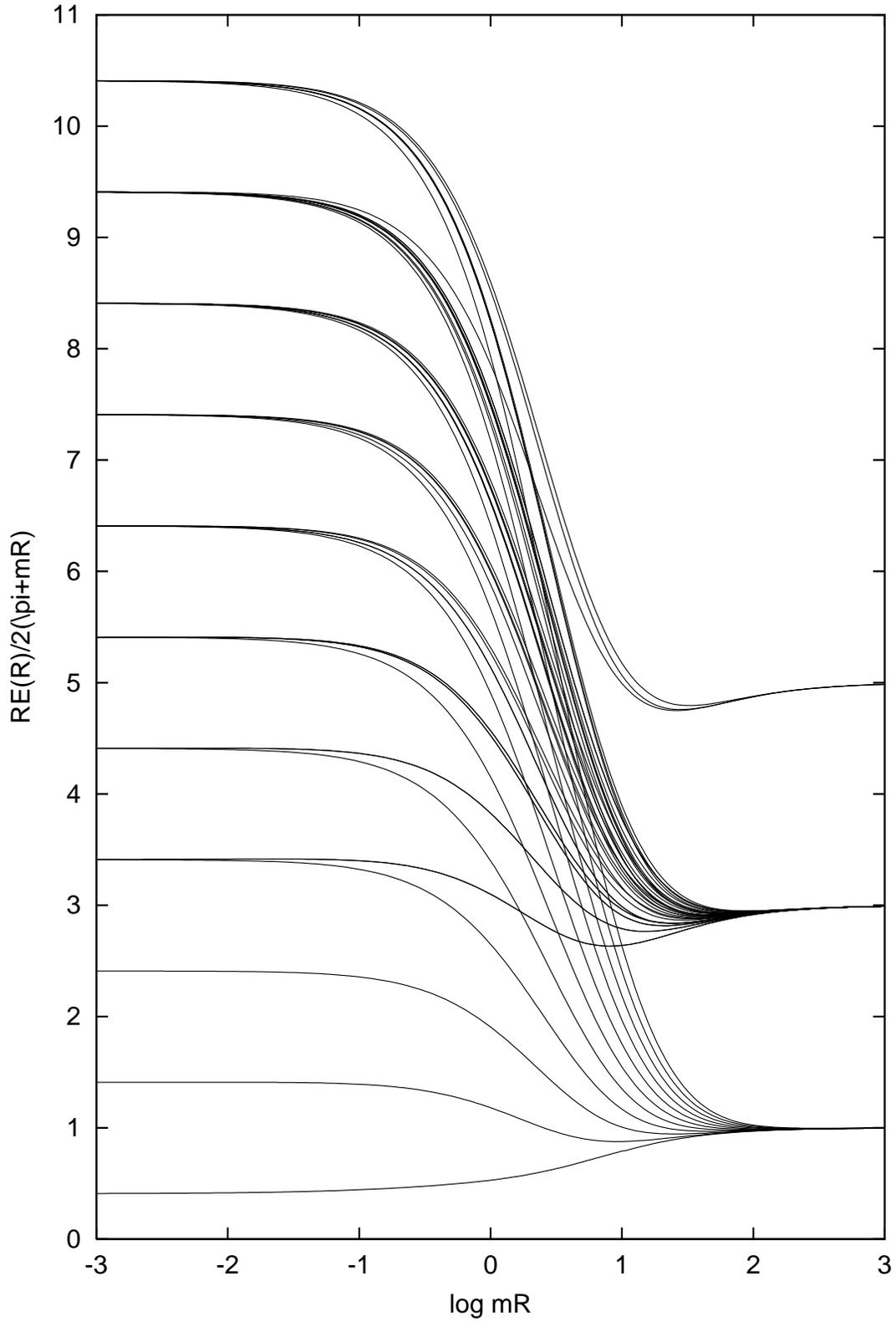}
\end{center}
\caption{Normalized scaling energies ${\cal E}(R)=\frac{RE(R)}{2(\pi+mR)}$ plotted against $\log_{10}
mR$ for the
$(r,s)=(2,1)$ boundary condition.}
\label{r2s1}
\end{figure}

In Figure~8 we show our numerical results for the $(r,s)=(2,1)$ boundary condition.
The vertical axis is the normalized scaling function ${\cal E}(R)$ and the 
horizontal axis is $\log_{10}(mR)$.  We plot selected normalized scaling energies for up to $m_1=5$
zeros in strip~1 and $m_2=3$ zeros in strip~2 for allowed quantum numbers including the lowest 18 
levels.  Table~3 summarizes how the UV descendant levels flow into the IR particle states.  The
plotted energy levels are complete  in the conformal limit up to $n=6$ corresponding to the expansion
of the finitized Virasoro character in the limit $N\to\infty$
\begin{eqnarray}
q^{c/24}\chi_{2,1}(q)&=&\qbin{\infty}{1}\qbin{1}{1}+q^3\qbin{\infty}{3}\qbin{2}{1}
+q^9\qbin{\infty}{5}\qbin{3}{3}
+q^{10}\qbin{\infty}{5}\qbin{3}{1}+\ldots\nonumber\\
&=&(1+q+q^2+q^3+q^4+q^5+q^6+q^7+q^8+q^9+q^{10}+\ldots)\nonumber\\
&&\quad\mbox{}+q^3(1+q+2q^2+3q^3+4q^4+5q^5+7q^6+8q^7+\ldots)(1+q)\nonumber\\
&&\quad\mbox{}+q^9(1+q+\ldots)+q^{10}(1+\ldots)+\ldots\\
&=&1+q+q^2+2q^3+3q^4+4q^5+6q^6+8q^7+10q^8+14q^9+18q^{10}+\ldots\nonumber
\end{eqnarray}

\section{Discussion}

In this paper we have derived and solved numerically the TBA equations for all excitations for the
massive tricritical Ising model with boundary conditions labelled by $(r,s)=(1,1)$, $(2,1)$ and
$(3,1)$. The analysis can be extended to the other primary boundary conditions $(r,s)=(2,1)$, $(2,2)$
and $(3,2)$  by allowing for frozen zeros and introducing two $(\vm,\vn)$ systems in the
classifications of eigenvalues. It would also be of interest to extend our analysis to periodic 
boundary conditions. The main new feature of such a calculation would be the classification of
the periodic eigenvalues which would entail many $(\vm,\vn)$ systems and must allow for different
patterns of zeros in the upper and lower half planes related to the two (left and right) copies of
the Virasoro algebra. If our analysis was extended to periodic boundary conditions it would
allow a direct comparison of the results of our lattice approach with the results of the Truncated
Conformal Space Approximation (TCSA) which are good for small $mR$.

In Part~II of this series of papers we will derive and solve numerically the TBA equations for all
excitations for the massless flow from the tricritical to critical Ising model. This has some
interesting additional features because zeros can collide during the flow leading to changes in the
classification of eigenvalues and to a flow between Virasoro characters of the two theories.

\section*{Acknowledgement}
PAP is supported by the Australian Research Council and thanks the Asia Pacific Center for
Theoretical Physics for support to visit Seoul.  CA is supported
in part by KOSEF 1999-2-112-001-5, MOST-99-N6-01-01-A-5 and thanks Melbourne University for
hospitality. PAP and CA thank Francesco Ravanini for hospitality at Bologna University where part of
this work was also carried out.

\bibliographystyle{amsplain}

\end{document}